\documentclass[11pt,aip,reprint,showpacs,preprintnumbers,amsmath,amssymb,groupedaddress]{revtex4-1}
\usepackage{graphicx}
\usepackage{bm}
\usepackage{fancyhdr}
\begin{document}
\title{Noise-induced synchronization in bidirectionally coupled Type-I neurons}
\author{Nishant Malik$^1$}
\author{B. Ashok$^2$}
%\email{basp@uohyd.ernet.in}
\author{J. Balakrishnan$^3$}
\altaffiliation{Corresponding author}
\email{janaki05@gmail.com}
\affiliation{
$^1$Potsdam Institute for Climate Impact Research, Telegrafenberg, 14412 Potsdam, Germany.}
\affiliation{$^2$Advanced Centre for Research in High Energy Materials (ACRHEM),\\
University of Hyderabad, Central University PO,
Gachi Bowli, Hyderabad - 500 046, India.}
\affiliation{ $^3$School of Physics, University of Hyderabad, Central University PO,
Gachi Bowli, Hyderabad - 500 046, India.}
%\date{\today}
\begin{abstract}
We present here some studies on noise-induced order and synchronous firing
in a system of bidirectionally coupled generic type-I neurons.
We find that transitions from unsynchronized to completely synchronized states occur
beyond a critical value of noise strength that has a clear functional dependence on
neuronal coupling strength and input values. For an inhibitory-excitatory (IE) synaptic
coupling, the approach to a partially synchronized state is shown to vary qualitatively
depending on whether the input is less or more than a critical value.
We find that introduction of noise can cause a delay in the bifurcation of the firing
pattern of the excitatory neuron for IE coupling.\\

{\Large{Published in Eur. Phys. J. B {\bf 74}, 177-193 (2010). DOI: 10.1140/epjb/e2010-00073-x }}
\end{abstract}
\pacs{05.45.Xt, ~05.45.-a, ~87.10.-e}
\maketitle
\vspace*{1cm}
{\bf Keywords:} ~~ Coupled type-I neurons, noise, complete synchronization, desynchronization\\
%\newpage
\section*{\large{1. Introduction}}
Among the host of interesting dynamical behaviour observed in coupled nonlinear
systems, synchronization phenomena are one of the most fascinating and among the most
studied. These phenomena abound in nature and in daily life: studies have been made of
synchronization in numerous systems --- in pendulum clocks, electronic
circuits, chemical systems, swarms of light-emitting fireflies, biological rhythms,
coupled Josephson junctions, cardiorespiratory interactions, neuronal ensembles in
sensory systems, etc. to name a few. Synchronous activity in neuronal ensembles
in particular, have received a great deal of attention and numerous studies
have been done on the spiking patterns of neurons and neuronal networks 
\cite{hodgkin,tuckwell,morrislecar,hanselsompolinsky,cerdeira,cerdeira2,rinzel,
vreeswijk,hansel,ermentrout,borgerskopell,borgerskopell2,izhikevich,lim,malik,
koch,gutkinetal}. 
Neuronal ensembles perform complex tasks and can extract different kinds of features 
from the information they receive and it is believed that cognitive tasks such as
feature extraction and recognition, sensory perception, etc., are brought about by
synchronous neuronal activity. The mechanisms and dynamics of synchronization phenomena
in neurons are therefore of great interest
\cite{rinzel,vreeswijk,hansel,ermentrout,borgerskopell,borgerskopell2,izhikevich,lim,
malik,koch,gutkinetal}.\\
Neurons are of many kinds (see for instance \cite{kandel}), but
they may be broadly classified into two types based on their excitability
mechanisms. This classification of neurons on the basis of the firing patterns
of the axons was first done by Hodgkin \& Huxley in their classic work \cite{hodgkin}.

Neurons of type-II  begin firing at a relatively high frequency through a subcritical
Hopf bifurcation.\\ 
Type-I neurons however can fire at arbitrarily low frequencies at low values of the 
applied input current, and the frequency increases with increasing values of the input. 
The equations which describe type-I dynamics can be reduced to the
canonical normal form for a saddle-node bifurcation \cite{rinzel}. In appropriate
parameter regimes the system operates in the close vicinity of a saddle-node
bifurcation on an invariant circle and this ensures repetitive firing.
It is interesting to study type-I neurons as almost all mammalian  neurons
fall under this class of excitability.\\ 

A variety of synchronous phenomena occur in nonlinear systems \cite{pikovsky}, but 
among these, complete synchronization (henceforth referred to as CS) is perhaps 
one of the most interesting \cite{pecora} 
as it brings about synchronization in the phases, frequencies and amplitudes of 
the signals. Although CS is widely believed to happen between identical systems, 
and CS between non-identical systems is not usually expected, we have shown 
in a different work \cite{malik} that in an ensemble of two hundred non-identical 
bidirectionally coupled type-I neurons with random strengths coupled in an 
all-to-all way, near complete synchronization does indeed unexpectedly occur 
under the influence of weak Gaussian white noise. 

In fact, a more extensive study of noise-induced synchronous activity
in a system of just two bidirectionally coupled identical (excitatory-excitatory)
type-I neurons reveals that the noise-induced CS state gives way beyond
a critical coupling and noise strength, to a desynchronized regime and then to subsequent
locking to a different partially synchronized state. We find that the value of
the critical coupling constant is proportional to the square root of the noise strength
\cite{malik}.
We were unable to achieve CS (noise-induced or otherwise) in a system of only two
bidirectionally coupled non-identical inhibitory-excitatory neurons, although we
showed in \cite{malik} that in a larger ensemble, non-identical neurons can exhibit
noise-induced CS. \\

In their studies of type-I neurons coupled through time-dependent synapses,
B{\"o}rgers and Kopell \cite{borgerskopell, borgerskopell2} discussed the
effects of random connectivity on synchronization
and the PING mechanism in networks of excitatory (E) and inhibitory (I) neurons. 

In this work, we report some interesting observations we have made in some
computer studies of coupled generic type-I neurons when subject to weak
additive Gaussian white noise. The dynamics of the synapses coupling the neurons
are governed by ordinary differential equations which depend upon the outputs of
the presynaptic neurons ~\cite{borgerskopell, borgerskopell2} --- we have
considered both EE and IE bidirectional couplings.\\
We find that the coupled system shows type-I firing pattern for EE coupling which 
suggests the existence of a contraction region close to the stable manifold of the 
saddle --- CS can thus be expected in EE coupled systems. This firing pattern is 
not seen for all the coupled neurons in IE systems --- the existence of a contraction 
region is therefore in question in this case, giving a good reason not to expect 
CS for IE couplings.

As pointed out in ~\cite{malik}, noise-assisted CS occurs in EE (identical) systems in
certain parameter regimes beyond which CS is destroyed but replaced by a
different partially synchronized state.\\ For such a system, we obtain here the
functional dependence between the critical value of the noise strength and the
strength of synaptic coupling and the externally applied input at the point when
desynchronization gives way to CS.\\
For the system of two coupled IE neurons, noise-assisted CS could not be achieved, 
but we find that
noise induces locking to a partially synchronized state at higher coupling strengths
and external input values. For the IE system, there exists a transition point at a
critical external-input value before which  desynchronization increases with
increasing noise strength, and beyond which desynchronization decreases with increasing
noise strength. \\ We also report here noise-induced delay in the bifurcation point
associated with the firing pattern of the excitatory neuron in the IE system.
We suggest that the fast-slow relaxation dynamics of the neuronal inputs in IE systems
make CS of the outputs hard to achieve.\\

We organize our presentation in the following way. In Section 2 we review the 
mathematical description of the mechanism of firing in type-I neurons and the 
framework for studying them when they are coupled. In Sections 3, 4 and 5 we describe 
our study of the coupled system in the presence of (weak) Gaussian white noise for 
EE and IE bidirectional couplings (Sections 4 and 5, respectively), separately.\\ 
For each type of coupling we make different studies to get insights upon the mechanisms 
governing synchronization. These are discussed in separate subsections: for an 
intuitive understanding of the dynamics underlying the approach to synchrony or order, 
in Section 4.1 we draw an analogy with the mechanical system of two damped nonlinear 
spring-mass systems coupled together. In Section 4.2, we describe our study of CS using 
plots between instantaneous frequencies and instanataneous phases of the inputs to 
the neurons and explain why these can be used as indicators of CS in coupled systems. 
In Section 4.3 we show how firing frequency-input plots for the coupled system can give 
indications of whether or not the coupled system is likely to show CS. The largest 
and second largest Lyapunov exponents have been calculated for the deterministic 
coupled system. 
In Section 4.4, we present our study of noise-induced variation of firing frequency. 
In the IE case, we show in the corresponding Subsection, the very interesting finding 
of noise-induced {\em delay} in the bifurcation associated with firing for the 
excitatory neuron. 
In Section 4.5, we present our study of variation of the synchronization error for 
CS, with noise strength, input, and coupling strength. We obtain useful functional 
relations between these for EE as well as for IE couplings. 
Results of similar studies are presented for IE coupling in Section 5 ; in 
Section 5.4 we report our very interesting observation of noise-induced delay of 
bifurcation associated with the neuronal firing.  
In Section 6, we show for EE coupling how the synchronization time varies with 
noise strength and suggest the possibility of exploiting the results for feature 
extraction. We conclude by summarizing the results of our work in Section 7. 
\section*{\large{2. Coupled theta neurons}}
In a type-I neuron, the mechanism of excitability comes about as follows 
\cite{rinzel},\cite{ermentrout}. For low
values of the control parameter (the input current), there is a stable fixed
point and an unstable fixed point which is a saddle. Trajectories forming
the unstable manifold leave the saddle point and enter the stable fixed point,
forming a loop in phase space. This closed loop therefore now contains the
stable and unstable fixed points. As the value of the control parameter is
increased, the two fixed points come closer together, and at the critical
value, they coalesce and then disappear, resulting in a stable periodic
solution corresponding to repetitive firing. 

The activity $x_i$ of the $i$th type-I neuron in an ensemble of $N$ neurons, can be
related to the membrane conductance and its dynamics can be described by:
\begin{equation}
\dot x_i = x^2 + I_i , ~~~~I_i = \beta_i + \sum_{j=1}^{N}\alpha_jg_{ji}s_{ji}~,
\end{equation}
where we have chosen to work in units in which the time constant for the membrane 
potential is set to unity \cite{rinzel},\cite{ermentrout}.
Its total input
$I_i$ comprises of the external input $\beta_i$ (which in this work we have taken
to be constant), and the contributions from the presynaptic neurons, with $i=1,\dots N$.
The synaptic gating variable $s_{ij}$ represents the fraction of
ion channels open in the $j$th presynaptic neuron, and its values always
lie in the range 0 to 1, reaching the maximal value when the neuron spikes.
$\alpha_j=+1$ denotes an excitatory synapse and $\alpha_j=-1$  models an
inhibitory synapse, while $g_{ji}$ denotes the measure of the
synaptic strength from neuron $j$ to neuron $i$; in our work we have considered $g_{ii}=0$.
Eqn.(1) has no fixed points for $I_i >0$.
Since any solution of the equation tends to infinity in a finite time, a nonlinear
transformation to new variables $\theta_i$ may be made \cite{ermentrout}:
~$x_i = \tan{\theta_i/2}$ which maps the real line onto a circle, and so avoids
this blow-up. In terms of the new variables, eqn.(1) becomes
\begin{equation}
\frac{d\theta_i}{dt} = (1 - \cos \theta_i) +
(\beta_i + \sum_{j=1}^{N}\alpha_jg_{ji}s_{ji}(\theta_j))(1 + \cos \theta_i)
\end{equation}
The point $\theta = \pi$ represents the point at infinity in eqn.(1) and is
interpreted as firing of a spike; this transformation gives the theta neuron its name.\\
The dynamics of the synapse $s_{ji}(t)$ follow from the differential equation
considered in \cite{borgerskopell,borgerskopell2,izhikevich} :
\begin{equation}
\frac{ds_{ji}}{ dt}= -\frac{s_{ji}}{\tau_{ji}} + e^{-\eta(1+\cos \theta_j)}
\frac{1-s_{ji}}{\tau_R}
\end{equation}
The synaptic decay time has been denoted by $\tau_{ji}$ and $\tau_R$ denotes
the synaptic rise time.\\
It is clear from eqn.(2) that the activity of the $i$th neuron is regulated by feedback
from $s_{ji}$  because  $s_{ji}$  depends upon $\theta_j$ which in turn
depends upon $\theta_i$ ~($j \ne i$) ~through the coupling term having $s_{ij}(\theta_i)$.
The coupled system under study is depicted in Fig.(1). 

Since the time dependence of the bifurcation parameter $I_i$ is brought about because of the
feedback  $\sum_{j=1}^{N}\alpha_jg_{ji}s_{ji}$ from the other neurons received through
the synapse, we chose in this study to keep the external input $\beta_i$ constant
throughout so that the dynamics governing the feedback could be understood better. 
\section*{\large{3. Coupled theta neurons in the presence of weak noise}}
At the cellular scale, the activity of real neurons can be influenced by a number
of other factors : physical as well as biochemical, and also by random influences:
for instance, fluctuations in the neurotransmitter levels at synapses, conductance
fluctuations in ion channels or thermal fluctuations \cite{koch}. It is therefore 
of great importance to study the influence of random noise on coupled neuronal dynamics.
In this work we have used Gaussian white noise $\xi(t)$ to model these random influences
on the dynamics of the coupled generic neurons in eqn.(2). We consider neurons 
which are coupled bidirectionally as shown in Fig.(1) and subject to Gaussian 
white noise $\xi(t)$ with the following properties:
$\langle \xi(t)\rangle = 0$ ~, ~$\langle \xi(t)\xi(t')\rangle = 2 \sigma \delta(t-t')$,
The stochastic variables are taken to obey Stratonovich calculus.
Addition of Gaussian white noise $\xi$ to eqn.(1), manifests as multiplicative noise
in eqn.(2) because of the change of variables to $\theta$, so that the equations now
take the form
\begin{equation}
\frac{d\theta_i}{dt} = (1 - \cos \theta_i) +
(\beta_i + \sum_{j=1}^{N}\alpha_jg_{ji}s_{ji}(\theta_j)+ \xi(t) )(1 + \cos \theta_i)
\end{equation}

We consider the neuronal output to be described by the variable
$u_i=(1-\cos \theta_i)/2$ (as in \cite{lim})
since its time series resembles that of the membrane potential in real neurons.
This transformation maps the resting point $x_i = 0$ corresponding
to $\theta_i =0$ to $u_i=0$, and the spiking point $\theta_i= \pi $ to $u_i=1$
via the relation $u_i = x_i^2/(1 + x_i^2)$.
This choice of variables enables us to get some new insights into the
dynamics underlying the onset of complete synchronization.
Eqns.(4) and (3) then become
\begin{eqnarray}
\dot{u_i}&=&2\Big(u_i+(\beta_i+\sum_{j=1}^{N}\alpha_jg_{ji}s_{ji} +\xi)(1-u_i)\Big)\sqrt{u_i(1-u_i)} \\
\dot{s}_{ji}&=&-\frac{s_{ji}}{\tau_{ji}}+\exp{(-2\eta(1-u_j))}\frac{(1-s_{ji})}{\tau_R}
\end{eqnarray} 

Among the different kinds of synchronous phenomena, we find CS most interesting 
because the trajectories of the mutually synchronizing systems become identical. 
The literature contains several papers on CS in different coupled systems; 
however systems with more complicated couplings, such as that described by eqns.(5)  
and (6) have not been intensively studied yet --- an adequately satisfying explanation 
of conditions and circumstances under which CS can occur in such systems is still 
lacking.\\ 
Noise-induced CS in systems through common additive white noise was studied in \cite{zhou} 
who showed that the existence of a significant contraction region in phase space 
was a necessary condition for occurrence of CS --- the systems studied in their 
work were the Lorenz and Rossler systems --- far easier to handle analytically in 
comparison to our system. 

For type-I neurons, we could expect CS among neurons through common noise alone, 
because of the existence of a contraction region close to the stable manifold of 
the saddle. When any two type-I neurons are coupled together as in 
eqns.(5),(6), then the nature of the eigenvalues of the stability matrix of the 
coupled system at the fixed points would determine whether such a contraction 
region could exist, and if so, its nature.\\  
However, for our system in eqns.(5),(6), linear stability analysis fails as the 
Jacobian becomes singular at the fixed points. Physically, this is a reflection 
of neuronal spiking with the inbuilt phase-resetting mechanism contained in eqns.(3). 
Moreover, the time dependence of the control parameter and its continuous dependence 
upon the feedback also makes an analytical approach non-straightforward.
We therefore perform some numerical and computer-based studies
which provide some useful insights on the coupled dynamics.  

Just as in \cite{zhou}, we define CS to take place between neurons 1 and 2 on 
obtaining a vanishing value for the 
quantity $\langle |u_1 - u_2| \rangle$ which is the synchronization error averaged 
over all iterations.   

It is well known that inhibitory couplings help in producing phase-synchronous 
activity in deterministic systems \cite{hansel,ermentrout,borgerskopell}. For the
system under study in Fig.(1) too, we found that
antiphase states are stable in the noiseless situation for EE synapses: these states
become completely in-phase in the presence of noise. This does not happen in general
for IE synapses \cite{ermentrout,malik}.  

In the uncoupled system neuronal firing starts after the threshold value $\beta=0$ 
is crossed.
In the coupled system under study (Fig.(1)), the picture changes and the input $I_i$
acquires time dependance; neuronal firing is then controlled by the different parameters
in eqns.(2) \& (3): the coupling strength $\alpha_jg_{ji}$, the decay time
of the synapse $\tau_{ji}$, the external input $\beta_i$, etc..\\ 

We proceed as follows in our investigation. We look at the behaviour of the neuronal 
output variables $u_1$, $u_2$ for different parameters in the system in the noiseless 
case. The addition of noise changes the behaviour, introducing order into the system. 
However, the outcome is different for the EE and IE cases. To avoid any confusion, 
we therefore detail our results separately, for EE and IE, respectively, in the 
coming subsections. A fruitful way of understanding the physics of any system is 
to make an analogy with a more common, familiar one. Rewriting equations (5) 
and (6) enables a resemblance of the coupled neurons to a nonlinear anharmonic 
oscillator to be observed, to help understand its behaviour. We do this below.  
\section*{\large{4. EE ~synaptic coupling}}
In Fig.(2), we depict for EE coupling, the variations of both the neuronal output 
variables $u_1$ and $u_2$ with the different parameters in the system for the 
noiseless case: the constant inputs $\beta_1=\beta_2$, the coupling strengths 
$g_{12}=g_{21}$, and the synaptic decay time $\tau_{ij}$, varying one at a time 
and keeping the other two fixed. These portrayals of the system's behaviour with 
changing parameters have been done differently, with the colour coding showing 
the varying values of the control parameter. 
The advantage of plotting them this way is essentially to portray the maximum possible 
information (for instance variation of both neuronal variables simultaneously) at a glance.  
The diagram distinguishes between states where both neurons are active and evolving 
(``firing'' in this context implying that there is a difference between consecutive 
outputs for both neuron 1 and neuron 2), and states where one or both neurons are quiescent 
or have reached a steady (unchanging) state (``non-firing'') implying that there is 
no difference between consecutive outputs of one or both neurons). 
Thus the following states would be classified as ``firing'' in the plot:\\ 
~(i)~$|u_i(t+1) - u_i(t)| > 0$ ~and ~~$|u_j(t+1) - u_j(t)| > 0$. ~$i\ne j$~;\\
~(ii)~$|u_i(t+1) - u_i(t)| = 0$ ~and ~~$|u_j(t+1) - u_j(t)| = 0$, ~$i\ne j$.\\
Thus, even the state where ~$u_1(t+1)=1$, $u_1(t)=1$, ~ with ~$|u_2(t+1) - u_2(t)| \ge 0$, ~
i.e., where neuron 1 has reached its peak value of 1 in consecutive firings, while 
the other neuron may have likewise peaked or have shown some or no change, we would 
classify it under the ``non-firing'' plot. Thus, it would be most appropriate to 
term these plots ``Activity Diagrams'', indicative of the state of active evolution 
for both neurons in the coupled system.\\  

In the noiseless case the neurons can be excitable and fire only if the inputs 
$\beta_i$ are above the threshold $\beta_i=0$. For $\beta_i<0$, there are stable 
and unstable fixed points, but no periodic firing can arise as they are below 
the threshold. The maximal value of unity for $u_i$ corresponds to the generation 
of the spike (peak value). 

In Fig.(2a), we observe in the first plot where $\beta_1=\beta_2$ is the varying  
parameter for fixed $g_{ij}$ and $\tau_{ij}$, that above the firing threshold $\beta>0$, 
the points in the plot for any $\beta$ trace out a quadrant of a circle --- this 
is especially apparent for large $\beta$, ($\beta\approx 4$) : this is a reflection of 
the stability of the antiphase states $u_1$ and $u_2$ in the EE case. In the adjacent 
(middle) plot where $g_{ij}=g_{ji}$ is the control parameter, all the points plotted are 
the square ``firing'' ones because the input has been chosen in this case to 
be $\beta_1=\beta_2=0.1 > 0$. Here again it was seen that the solutions $u_1$ and 
$u_2$ are out of phase and the points trace out a roughly circular curve: The last 
plot in Fig.(2a) where the decay time $\tau_{ij}=\tau_{ji}$ of the gating variable 
is the parameter shows that the magnitudes of $u_1$ and $u_2$ do not show 
appreciable change with change in magnitude of $\tau_{ij}$.\\ 
To make the plots clearer, we have plotted the variation of both neurons $u_i$ with 
$\beta$ in the standard manner, in Fig.(2b). For $\beta<0$ the neurons do not fire 
and each solution is non-oscillatory; after the threshold is 
crossed however, i.e., for $\beta>0$, both neurons show oscillatory behaviour 
--- the plot therefore shows bifurcation of neural activity. 
We observe that $u_1$ and $u_2$ oscillate out of phase with each other. 

In Fig.(3) we show the development of noise-induced synchrony as the noise strength 
is increased from 0.375 (left) to 1.0 (right) for EE coupling: a drastic change in 
the firing activity pattern is observed on introduction of weak noise: in contrast to 
Fig.(2), all neurons now fire. Even for negative values of $\beta_1$ \& $\beta_2$ 
(corresponding to stable points in the flow of the uncoupled system: the neurons 
do not fire in the deterministic system), addition of weak noise induces firing 
by raising total input levels to threshold values. One observes that the nearly 
symmetrical distribution of points on either side of the diagonal $u_1- u_2=0$~ 
in (Figs.(2a)) in the noiseless case, converges towards the diagonal in the 
presence of noise. As the noise strength is increased, the firing of the two 
neurons is completely synchronized  and for $\sigma=1.0$, all the points line up 
along the diagonal and are CS solutions. 
\subsection*{4.1 Analogy to a mechanical system}
We can get some insight into the dynamics underlying the development of CS in 
the EE systems by drawing an analogy with a mechanical system. To do this, we rewrite 
eqns.(5),(6) for $N=2$, $\alpha_1=1$, $\alpha_2=1$ as: 
\begin{eqnarray}
\dot u_1 &=& f_D(u_1) +\beta_1f_n(u_1) +g_{21}(s_{21}+1)(u_2)f_n(u_1) +\xi f_n(u_1) 
\nonumber\\
\dot u_2 &=& f_D(u_2) +\beta_2f_n(u_2) +g_{12}(s_{12}+1)(u_1)f_n(u_2) +\xi f_n(u_2) \nonumber\\
\ddot s_{21} &+& h(u_2)\dot s_{21} +(\Phi(u_2) +\Psi(u_2)g_{12}(s_{12}+1))s_{21} \nonumber \\ &=& -\Psi(u_2)s_{21}\xi(t) \nonumber\\
\ddot s_{12} &+& h(u_1)\dot s_{12} +(\Phi(u_1) +\Psi(u_1)g_{21}(s_{21}+1))s_{12} \nonumber \\ &=& -\Psi(u_1)s_{12}\xi(t)
\end{eqnarray}
where
\begin{eqnarray}
f_D(u_i)&=& 2u_i^{3/2}{(1-u_i)}^{1/2} ~~;~~ f_n(u_i)=2u_i^{1/2}{(1-u_i)}^{3/2} ~;~\nonumber\\
h(u_i)&=& \Big(\frac{1}{\tau_{ij}}+\frac{e^{-2\eta(1-u_i)}}{\tau_R}\Big) ~~; \nonumber\\
\Phi(u_i)&=& 2\eta\frac{e^{-2\eta(1-u_i)}}{\tau_R}(f_D(u_i)+\beta_if_n(u_i)) ~~;~\nonumber\\
\Psi(u_i)&=& 2\eta\frac{e^{-2\eta(1-u_i)}}{\tau_R}f_n(u_i) 
\end{eqnarray}
and where a shift of $s_{ij}$ by unity has been performed in eqns.(7) 
(and later on in eqns.(14)) for the sake of putting these 
in the more convenient form displayed. 
As can be seen, these equations are not easily amenable to a quick analysis of the physics
that they describe.
The equations for the gating variables  $s_{ij}$ in eqns.(7) resemble those of a nonlinear
anharmonic oscillator with a nonlinear damping term and a nonlinear restoring force term
with variable spring coefficient ~$(\Phi(u_i) \pm \Psi(u_i)g_{ji}(s_{ji}+1))$,
with the plus sign for excitatory couplings and minus sign for inhibitory couplings. 
In the EE system, the variable spring coefficient is ~$(\Phi(u_i) + \Psi(u_i)g_{ji}s_{ji})$
for both $s_{21}$ and $s_{12}$ and it is like a system of two damped (nonlinear) spring-mass
systems coupled together, with both ``springs'' getting extended or both getting compressed
(though by different amounts) at the same time. This, coupled with the same (plus) sign in
the $u_i$ equations lead to both neurons eventually firing at the same rate if the neurons
happen to be identical.\\
The addition of common weak noise $\xi(t)$ to both neurons (eqns.(5))
changes the potential at the synapses (in mechanistic terms, it affects the ``stiffness''
of the synaptic conductances in eqns.(7)), and also modulates the firing thresholds
of $u_1$ and $u_2$ by injecting energy at random points in time.\\
For EE coupling, the thresholds of each neuron is modulated similar to the other due 
to the common noise and if the noise strength is adequate, neuronal firing 
eventually occurs in synchrony. 
\subsection*{4.2  Variation of instantaneous frequency with instantaneous phase of neuronal inputs}
To understand how this CS actually comes about, in \cite{malik} we constructed the 
analytical signal
$B_i(t)e^{i\rho_i(t)}= I_i(t) + iH(I_i(t))$ for
the inputs $I_i$ and those for the neuron outputs $u_i$ : $w_i(t)=u_i(t) + H(u_i(t))
= R_i(t)e^{i\phi_i(t)}$ using Hilbert transforms and showed that the instantaneous amplitudes
$B_i$ and instantaneous phases $\rho_i$ of the inputs evolve in time according to:
\begin{eqnarray}
\dot B_i(t) &=& -B_i\Big(\frac{1}{\tau_{ji}}+\frac{e^{-2\eta(1-R_j\cos\phi_j)}}
{\tau_R}\cos(2\eta R_j\sin\phi_j)\Big) \nonumber\\ &+& \frac{\beta_i}{\tau_{ji}}\cos\rho_i \nonumber\\
&+&\frac{1}{\tau_R}\Big(\beta_i+\sum_j\frac{\alpha_jg_{ji}}{\tau_R}\Big)
\cos\Big((2\eta R_j\sin\phi_j)-\rho_i\Big)\nonumber\\
&\times& e^{-2\eta(1-R_j\cos\phi_j)}\nonumber\\
\dot\rho_i(t) &=& -\frac{e^{-2\eta(1-R_j\cos\phi_j)}}{\tau_R}\sin(2\eta R_j\sin\phi_j)
-\frac{\beta_i}{B_i\tau_{ji}}\sin\rho_i \nonumber\\
&+&\frac{1}{\tau_R B_i}
\Big(\beta_i+\sum_j\frac{\alpha_jg_{ji}}{\tau_R}\Big)\sin\Big((2\eta R_j\sin\phi_j)-\rho_i\Big)\nonumber\\
&\times&e^{-2\eta(1-R_j\cos\phi_j)}
\end{eqnarray}
We argued in \cite{malik} that CS is achieved if the rate of change of {\em instantaneous}
frequency ($\dot\phi$) with the {\em instantaneous} phase ($\phi$) of the input for
neuron 1 matches exactly that for neuron 2. This is not a trivial statement since
the system incorporates feedback as well as noise.\\ 
This matching of values happens because noise produces delay in the decay time of 
the synaptic gating variables, causing CS to eventually occur. 
In Fig.(4a) we demonstrate this through a sequence of time series of $s_{ij}$ for 
increasing noise strengths. 
We find that addition of weak noise delays the decay time 
of $s_{ij}$ gradually, eventually lowering its minimum to zero. This produces a 
delay in the onset of the next $s_{ij}$ peaks which in turn delays the input $I_i$ to 
the post-synaptic neuron at a future time instant. The decay time of $s_{ij}$ is 
increased further as the noise strength is increased. The periodically occurring 
maximal values of the inputs then arrive at the post-synaptic neurons later than in 
the previous cases and this gets reflected in the firing pattern of neurons as 
departures from the noiseless /previous values of the differences $(u_1-u_2)$ and 
those of the phase differences between the neurons.
This can be seen more clearly in Fig.(4b) where the synaptic conductance for a single 
neuron is shown for increasing noise strengths.  

For a given set of $\beta$, $\tau_{ij}$, $\eta$, $\tau_R$, the values of the 
instantaneous phases $\phi_i$ and those of their temporal variation, i.e., the 
instantaneous frequencies, are determined by the synaptic coupling strengths 
$\alpha_{ij}g_{ij}$ and the noise strength, which are inputs received at the synapse. 
When the variation of the instantaneous values with instantaneous phases of the 
{\em inputs} to neuron 1 matches with that for neuron 2, then it is reasonable 
to expect CS to occur in the EE system, because the instantaneous values $\phi_1$ 
and $\dot\phi_1$ of neuron 1 change in step with $\phi_2$ and $\dot\phi_2$ 
respectively of neuron 2. This forces the amplitudes of neurons 1 and 2 to become 
identically the same, because if they do not, then both conditions $\phi_1=\phi_2$ 
and $\dot\phi_1=\dot\phi_2$ cannot simultaneously be maintained. 
Such a plot (instantaneous frequency $\dot\phi$ versus instantaneous phase 
$\phi$) of the {\em input} to each neuron then has a characteristic flame shape
(Fig.(5)), which exactly matches with that for the other neuron when CS occurs
\cite{malik}, and this identity of the plots can be used as an indicator of CS. 
\subsection*{4.3  Input-Frequency curves and Lyapunov exponents}
The input-firing frequency curves for the coupled system plotted in Fig.(6) for 
different values of the coupling strength are typical characterizations of type-I 
neurons, differing from them only in that, even at $\beta_i=0.0$ there is a non-zero 
frequency of neuronal firing, since $I_i \ne 0$. Lyapunov exponents have been 
calculated and shown alongside. 

These frequency-input curves indicate firing via a saddle-node bifurcation for 
the coupled EE system. 
Since the occurrence of a saddle-node bifurcation allows for a contraction region along 
and close to the stable manifold of the saddle, and in conformity with the result 
of \cite{zhou}, it seems reasonable to expect CS to occur among systems coupled 
through EE synapses.\\ 

Fig.(6a) depicts identical neurons with EE coupling
($g_{ij}=g_{ji}=0.3$ and other parameters as shown): the system is periodic
with the presence of only a single frequency. Increasing the coupling strength of one
neuron to $g_{ji}=0.45$, all other parameters and conditions remaining as before
(corresponding to non-identical neurons with EE coupling) produces
a second frequency after $\beta_i$ crosses a value of approximately $0.7$ (Fig.(6b)).
This is manifested in the Lyapunov exponent curve: the second largest Lyapunov exponent
also becomes zero at the same point, indicating the presence of a second frequency
in the system. As this transition from a periodic to a quasiperiodic state involves
variation in $g_{ji}$ and $\beta_i$, for given $\tau_{ij}$, $\tau_R$, a codimension
two bifurcation is indicated at this point.
\subsection*{4.4 Noise-induced variation of firing frequency}
Largest Lyapunov exponents in the presence of noise were calculated in \cite{malik}
using the stochastic Runge-Kutta-4 method.
We found that in both EE and IE cases, the largest Lyapunov exponent $\lambda_1$
becomes more negative on the addition of noise demonstrating the role of noise
in bringing about order into the system.\\ 
In the presence of weak noise, firing frequency changes with $\beta_i$
differently for the excitatory and the inhibitory neurons, depending upon whether
the coupling is of EE or IE type. For EE coupling, this is displayed in Fig.(7) 
and the plots for the noiseless case is shown in red in Fig.(7(b)).  

One observes that the frequency of firing of both excitatory neurons
starts from a large non-zero value even at $\beta_1=\beta_2=0$ (Fig.(7)).
As $\beta_i$ is increased further, the frequency decreases rapidly, and after a certain
value of $\beta_i$, the curve merges with that for $\sigma=0$, taking on the very
same frequency values as for the noiseless case, so that from this point onwards,
the frequency once again increases with increase of $\beta_i$, but the increase is now
gradual.
\subsection*{4.5 Noise-induced synchronization \& locking to partially synchronized state}
Numerous studies on CS in coupled systems have been previously reported in other systems
(see for instance \cite{hanselsompolinsky,cerdeira,cerdeira2,pikovsky,pecora,
zhou,anishchenko} and references therein).
Noise-induced CS in coupled theta neurons happens for EE coupling in a certain
parameter range.  However, we have not yet succeeded in observing it for IE coupling
in a system of just two neurons.
There appears to be definite relations between the magnitude of the output-difference
between the two neurons averaged over all iterations $\langle|u_1-u_2|\rangle$,
and the various parameters of the theory such as $g_{ij}$, $\beta_i$,
and the noise strength $\sigma$, for both EE \& IE couplings, and these may be quantified
by plotting these quantities (Figs.(8)-(13)).\\
In Fig.(8), $\langle|u_1-u_2|\rangle$ is plotted as a function of $g_{ij}$ for
$\beta_1=\beta_2=0.1$ in the case of EE coupling. It is seen that increasing 
the noise strength helps in bringing about synchronization for low values of 
$g_{ij}$. At higher coupling strengths, CS is destroyed as $\langle|u_1-u_2|\rangle$ 
increases steadily, but then it plateaus off, approaching a limiting value of 
approximately 0.5 at about $g_{ij}=6$ for all noise strengths.
Partial synchronization therefore takes place as the system gets locked to this state.\\ 
In \cite{malik} we found for EE synapses that for a given input, the critical 
coupling strength $g_c$ before which the neurons are completely synchronized, and 
beyond which they are desynchronized depends upon the noise strength as:
\begin{equation}
g_c \sim a\sigma^{1/2}
\end{equation}
where for the choice of parameters we had considered, we had $a \approx 1.1$. In that work,
we also found for curves such as those in  Fig.(8), the following functional dependence
of the synchronization error on $g$ for the entire regime of coupling strengths 
after the onset of desynchronization:
\begin{equation}
\langle |u_1 - u_2| \rangle \sim a(\sigma) - b(\sigma) - \frac{g_c^4}{g^3}
\end{equation}
where the constants $a$ and $b$ are noise-dependent.\\

The variation of $\langle|u_1-u_2|\rangle$ with $\beta_i$ for EE synapses can be 
seen in Fig.(9).
CS does not happen for $\sigma=0$, but increasing noise strength decreases the difference.
At lower nonzero noise strengths, the system oscillates between the completely synchronized
and the unsynchronized states. At a moderate noise strength of $\sigma=0.6$, the system
shows CS for nearly the entire range of $\beta_i$ shown in the figure. 

In Fig.(10), we show the variation of $\langle|u_1-u_2|\rangle$ with noise 
strength at different $g_{ij}$ values, for $\beta_i=0.1$. We find that 
in neurons coupled through EE synapses, the transition to CS occurs at larger noise 
strengths for stronger coupling strengths; for weaker couplings, smaller noise 
strengths suffice for CS to occur. The system makes large oscillations between two 
metastable unsynchronized states before settling into the stable synchronized state. 
We find from an analysis of the plot in Fig.(10) that  the point where desynchronization 
gives way to synchronization, i.e., $\langle |u_1 - u_2| \rangle \rightarrow 0$  
begins to be approached, can be given by the relation
\begin{equation}
\sigma_{turn} \approx  0.5 ~g^{3/4} + \beta
\end{equation}
We find that the maximum value that $\langle |u_1 - u_2| \rangle$ takes is at minimal
noise strength $\sigma$ and is related to the coupling strength $g$  (for $\beta=0.1$) by
\begin{equation}
\langle |u_1 - u_2| \rangle_{max} \approx  0.607 ~g^{0.1}
\end{equation}
This relation is made clear from the plot in fig.(10c).\\ 
We now discuss our results for neurons coupled through IE synapses. 
\section*{\large{5.  IE ~synaptic coupling}}
In Fig.(11a) we depict, as for the EE case, variations of both neuronal variables 
$u_1$ and $u_2$ simultaneously, with $\beta$, $g_{ij}$ and $\tau_{ij}$, 
in the absence of noise where the out of phase solutions are not the stable ones. 
The ``firing'' (active) points are more randomly distributed in this case.\\  
In contrast with the EE case, for IE coupling, we observe in Fig.(11a) that the 
distribution of the solutions about the diagonal $u_1- u_2=0$~ is far from 
being symmetric. Fig.(11b) shows the variation of $u_i$ activity with $\beta$ in 
a more standard way. The solutions for $\beta$ less than the threshold value 
(ie for $\beta<0$) show stable, non-oscillatory behaviour. All $u_i$ for $\beta>0$ 
however show large oscillations.  

In the presence of noise, the IE system never achieves CS of its solutions.  
Introduction of weak noise does however bring about some order even in the IE case
and this can be visualised in the plots in Figs.(12) for noise strengths 0.5 and 1.0,
where consecutive points in the $u_1$-$u_2$ plane have been connected through a curve.
The originally zig-zag trajectories in the noiseless case are now
replaced by smooth curves which evolve in time exploring large regions of the phase 
space for increasing values of $\beta_i$ in a definite, orderly manner. 
\subsection*{5.1 Analogy to a mechanical system}
In the case of IE synaptic couplings, addition of even very weak noise greatly 
affects the neuronal outputs, firing rates and the inputs they receive.
For this type of coupling, it is clear from the time series that the inputs
have fast and slow branches always, characteristic of relaxation dynamics. 
In this case, eqns.(5) and (6) can be rewritten (similar to the EE system) 
for $N=2$, and $\alpha_1=1, \alpha_2=-1$, as:
\begin{eqnarray}
\dot u_1 &=& f_D(u_1) + \beta_1f_n(u_1) - g_{21}(s_{21}+1)(u_2)f_n(u_1) + \xi f_n(u_1) \nonumber\\
\dot u_2 &=& f_D(u_2) + \beta_2f_n(u_2) + g_{12}(s_{12}+1)(u_1)f_n(u_2) + \xi f_n(u_2) \nonumber\\
\ddot s_{21} &+& h(u_2)\dot s_{21} + (\Phi(u_2) + \Psi(u_2)g_{12}(s_{12}+1))s_{21} \nonumber\\ &=& -\Psi(u_2)s_{21}\xi(t) \nonumber\\
\ddot s_{12} &+& h(u_1)\dot s_{12} + (\Phi(u_1) - \Psi(u_1)g_{21}(s_{21}+1))s_{12} \nonumber\\ &=& -\Psi(u_1)s_{12}\xi(t)
\end{eqnarray} 
Here the nonlinear restoring force has variable spring coefficient 
$(\Phi(u_1) - \Psi(u_1)g_{21}(s_{21}+1))$. This restoring force is incremented
for $s_{21}$ but reduces  for $s_{12}$. The system can be visualised crudely as
two damped spring-mass systems, one with a very stiff spring and the other with a very
loose spring coupled together. The extension of one and simultaneous compression of the other
spring leads to separation of time scales and relaxation oscillations of the synaptic conductances
(Fig.(13)). These are fed as inputs to the neurons, again in the $u_2$ equation with a plus sign,
while in the equation for $u_1$ with a minus sign. This discordance in the inputs results
in non-synchrony of the neuronal outputs, and the excitatory neuron fires at a frequency
different from the inhibitory neuron; this mutual difference in firing frequency
is maintained even on increasing the common external constant input $\beta$: this is 
discussed further on in Section 5.3. 
Indeed, a canard solution (Fig.(13b)) appears to govern the underlying dynamics. 
\subsection*{5.2 Variation of instantaneous frequency with instantaneous phase of neuronal inputs} 
For IE coupling, though modulation of the firing thresholds by common noise
occurs simultaneously for the two neurons, the mutual discordance of the inputs near 
the firing thresholds because of the opposite signs in the respective coupling terms 
in eqns.(14) make it hard for the two neurons to cross the threshold and fire at the 
same points in time, as we discussed above, using the mechanical analogy with the 
coupled damped spring-mass systems.
Therefore although noise induces order even in IE systems (see Fig.(12)), yet complete
synchronization becomes very hard to achieve because of the relaxation dynamics of 
the inputs: the ``flame plots'' for the two neurons do not match in this case 
(see Fig.(14) for example). 
\subsection*{5.3 Input-Frequency curves and Lyapunov exponents} 
Unlike the EE case, we cannot expect CS in the IE case, because the firing pattern 
(Fig.(15)) does not indicate presence of a saddle for both neurons --- the firing 
frequency of the excitatory neuron shows non-monotonic behaviour with respect to 
the constant input $\beta$: exhibiting very high firing frequencies at $\beta\approx 0$, 
then a sudden decrease at $\beta\approx 0.3$ and then a regular monotonic increase 
thereafter, and tending to maintain a nearly constant frequency difference with 
the inhibitory neuron for larger $\beta$ in the deterministic system where $\sigma=0$. 
The nearly constant frequency difference however is suggestive of order emerging 
in their phase differences with increasing $\beta$. \\
The plots in Fig.(15) are for parameter values identical with those in Fig.(6a) except
for the coupling which is of IE type, (so that we now have now non-identical neurons): 
here again two different frequencies are present. In this case, for small $\beta_i$ 
values, the large fluctuations in the frequency of the excitatory
neuron (Fig.(15a)) when $\sigma=0$ arise because of the sign changes of the input ~
$(\beta_i + \sum_{j=1}^{N}\alpha_jg_{ji}s_{ji}(\theta_j))$ since $\alpha_j=-1$ ;
the bifurcation parameter oscillates between the stable region where $I_i<0$ and
the unstable region $I_i>0$.
Similar fluctuations are observed in the Lyapunov exponent curve also (Fig.(15b)), 
as shown in the inset curve, and the largest exponent even becomes positive in a small
window of neuronal inputs. 
\subsection*{5.4 Noise-induced variation of firing frequency \& delay of bifurcation}
For IE coupling, addition of noise causes the firing frequency of the
inhibitory neuron to start from a high non-zero value just as in the EE case
as the sign of the input term remains unchanged. However, for the excitatory neuron
of Fig.(15), there is a non-monotonic dependence of its firing frequency on
$\beta_i$  (see Figs.(15),(16b)).
In the absence of noise, when $\beta_i=0$ and $\alpha_j=-1$ then $I_i<0$  and the
neuron is below the firing threshold; hence the excitatory neuron does not fire.
Introduction of weak noise however injects the required energy to the neuron to
overcome the threshold and fire. In Fig.(16) we show that the frequency of firing in
the presence of noise is higher than in the noiseless case (shown in red) but 
as the external input $\beta$ is increased, the effect of noise on the firing 
frequency dies down. \\

We observe a very interesting phenomenon in the IE case when noise is introduced:
the irregular firing of the excitatory neuron for small $\beta_i$ values which was
accompanied by two sharp peaks in the frequency (at the small $\beta_i$ values
of  0.01 \& 0.1) in the absence of noise, is replaced in the presence of noise by
a single broad peak at the much larger values of $\beta_i=0.65$ for $\sigma=0.025$,
and $\beta_i=1.6$ for $\sigma=0.075$ (Fig.(16b)). Here noise appears to
{\em delay} the bifurcation associated with this different firing behaviour.\\
It is well known that the sweep of a control parameter which is slowly varying in time,
causes a delay in the steady bifurcation point \cite{haberman}. The effect of
Gaussian white noise into such a system reduces this delay in the bifurcation point
depending upon whether the noise is additive or multiplicative \cite{vandenbroeck}.
In eqns.(5) and (6) Gaussian white noise enters the system multiplicatively. The
neuronal inputs especially for the IE synapse show relaxation oscillations, having
slow and fast time scales. This is quite distinct from the cases studied in
\cite{vandenbroeck}, where inputs vary very gradually and slowly with time.
We note that the delay in the bifurcation associated with the change in the firing
behaviour is for the {\em excitatory} neuron of the IE system, and not for the
inhibitory neuron.  

As mentioned earlier, we observe from the time series that noise causes a delay
in the synaptic decay time \cite{malik}. While in the EE case this results
in a change in the onset of the arrival of the maximal input values to the neurons
and eventual synchronization (CS) of the neuronal outputs \cite{malik} , in
the IE case, because the signs of the input terms of the two neurons are mutually
opposite, the neurons eventually are never able to fire in-step and
hence CS is almost never achieved.\\

Noise-induced delay of bifurcation in coupled theta neurons has been
reported earlier in the literature \cite{gutkinetal}, but the coupling in their
case was purely of EE type. 
\subsection*{5.5  Noise-induced synchronization \& locking to partially synchronized state}
For IE coupling (Fig.(17)), the synchronization error $\langle|u_1-u_2|\rangle$ has 
a maximum magnitude at $g_{ij}=2.4$ approximately, for all values of noise strengths, 
and approaches a saturating value of 0.27 at about $g_{ij}=6$ irrespective of the 
noise strength, both for $\beta_i=0$ and for $\beta_i=0.1$.  
There are three interesting things to be noted: first, values of  
$\langle|u_1-u_2|\rangle$  are lower in the IE case in comparison to EE coupling, 
but in the IE case too there is partial synchronization for large coupling strengths.
The second point is that increasing noise strength beyond a critical value $g_c=0.4$
increases the difference between neuron outputs while below this value the difference
decreases for nonzero $\sigma$, in contrast to the EE case where increasing noise strength
brought down their difference. Finally, we see that for IE coupling in the noiseless situation,
the variation of $\langle|u_1-u_2|\rangle$  with  $g_{ij}$ at $\beta_i= 0.1$ is
discontinuous at a small coupling strength of about $g_{ij}=0.3$ (Fig.(17)). 
The reason for the discontinuity is not yet clear to us.\\ 
Variation of $\langle |u_1 - u_2| \rangle$  with the input $\beta$  for the IE case
with $\tau_1=\tau_2$, $\beta_1=\beta_2=\beta$, $g_{12}=g_{21}$ is shown in Fig.(18) for
different noise strengths. For the chosen values of parameters, we observe that
there is a transition point at $\beta = \beta_c \approx 0.1$.
For $\beta > \beta_c$, $\langle |u_1 - u_2| \rangle$ decreases with increasing $\sigma$. 
There is a cross-over region between $\beta=\beta_{c_{l}}=0.03$ and $\beta_c$, and 
for $\beta < \beta_{c_{l}}$, $\langle |u_1 - u_2| \rangle$ increases with increasing 
$\sigma$. The plots show a dependence of the form:
\begin{equation}
\langle |u_1 - u_2| \rangle \sim a \beta^b,   ~{\rm for}~  \beta \leq \beta_{c_l}
\end{equation}
where $a$ \& $b$ vary with the noise strength $\sigma$  and
\begin{equation}
\langle |u_1 - u_2| \rangle \sim k\beta + l\beta^{1/2}
+ (2-\sigma) \beta^{2/5} ,   ~{\rm for}~  \beta > \beta_c
\end{equation}
where $k,l$ vary from curve to curve and depend on the noise strength $\sigma$.\\
However, as mentioned earlier also, $\langle|u_1-u_2|\rangle$ never becomes zero for IE synapses;
instead, at large $\beta_i$ values, the output difference saturates for all noise strengths
to a limiting value of about 0.325 and the system gets locked on to this state.
The discontinuity at $\beta_i= 0.1$ for the noiseless case (in red) is a reflection of
very small, non-negative values of the Lyapunov exponents (Fig.(15b), inset)
(discussed also in the previous section) because of the total input to
the excitatory neuron oscillating between the stable regime when $I_i<0$ and
the unstable regime when $I_i>0$, depending upon the relative strengths of $\beta_i$ and
$\alpha_jg_{ji}s_{ji}$, since $\alpha_j=-1$ for neuron $j$. 

The variation of $\langle|u_1-u_2|\rangle$ with noise strength for IE synapses  
at different $g_{ij}$ values, for $\beta_i=0.1$ is displayed Fig.(19). 
Above a noise strength of about $\sigma=0.15$, increasing $g_{ij}$ increases 
the difference between the neuron outputs; below $\sigma=0.15$ also the system 
remains far from CS, but increasing $g_{ij}$ can bring down the difference. 
Comparing these IE plots with the ones in Fig.(10) for EE, we see that as usual EE 
and IE synapses show different behaviour.
The uncoupled case $g_{ij}=0$ is shown in both plots of Figs.(19),(10) for comparison. 
\section*{\large{6. Variation of synchronization time with noise strength (EE coupling)}} 
In Fig.(20) we plot the time taken by the EE system to exhibit CS as a function of the noise
strength, keeping all other parameters fixed. The normal clock function available with
C-programming language has been used to measure the time $T_{syn}$ taken by the system to
synchronize.\\ 
To define $T_{syn}$, the procedure we use is as follows. For a given value of the noise 
strength $\sigma$, the absolute value of the output difference $|u_1-u_2|$ is calculated,  
at each instance, for a large number $k$ of iterations ($k = 300000$). Once $|u_1-u_2|$ 
goes to zero in, say the $i$th iteration, (or more accurately, when this difference can 
be considered to be negligible computationally --- less than $10^{-6}$), 
the system is considered to have synchronized only if for all the consecutive and 
remaining $(k-i)$ iterations, $|u_1-u_2| \approx 0$  continues to hold good. 
When this synchronization condition is found to be satisfied, then the time 
(from the beginning of computation) at which $|u_1-u_2| \approx 0$ first occurred 
(at the $i$th iteration), is considered to be the synchronization time $T_{syn}$.  

We observe that $T_{syn}$ is smaller for higher noise strengths.
Infinitesimally small noise strengths suffice to completely synchronize two uncoupled
neurons ($g_{ij}=0$), though the time taken for this to happen may be longer.
In this case, increasing noise strength causes CS to occur for very small values of $T_{syn}$.
For increasing coupling strengths, there exist plateaus or bands of
$\sigma$ for which $T_{syn}$ is the same. For stronger couplings ($g_{12}=g_{21}=0.5$
and $g_{12}=g_{21}=1.0$) we observe the existence of metastable states in certain regimes
of noise strengths where the times taken for CS to occur can be different.\\
This suggests the possibility of coupled neurons utilizing the different synchronization
times for encoding different features from the given inputs. Indeed, this opens up the
possibility of exploiting ``$T_{syn}$ switches'' in artificial neural networks for feature
extraction and recognition. 
\section*{\large{7. Conclusion}}
We have reported in this work results of some computer studies on noise-induced
complete synchronization in systems of bidirectionally coupled type-I neurons.
We have studied both EE and IE couplings.
From analyses of these results, we obtain functional relations between the
input strength, noise strength, coupling strengths and the distance from CS.
Transitions from a completely synchronized regime to a partially synchronized state
to which the system gets locked with increasing coupling strength, or with increasing
input, other parameters kept constant, are observed. Similarly we observe that with
increasing noise strength desynchronization yields to CS. We find that the critical
noise strength where this occurs has a clear functional relationship to the
coupling strength and the input. For IE coupling, we find that depending on whether the
input $\beta$ is less or more than a critical value $\beta_c$, the system's departure
from completely synchronized to a partially synchronized state occurs very differently.

We find that Gaussian white noise can cause a delay in the bifurcation
associated with the firing of the excitatory neuron when the synapse is of IE type.
We suggest that relaxation dynamics of the neuronal inputs in
IE systems make CS of the outputs hard to achieve even in the presence of noise.
Finally, we point out the possibility of utilizing the different times for
complete synchronization as switches for extracting different features in
networks of neurons.   
\section*{\large{8. Acknowledgements}}
We would like to thank the anonymous referees for valuable comments and  
useful suggestions. 

\newpage
\noindent
{\bf Figure Captions}\\

{\bf Figure 1.} (color online) Bidirectional coupling between two theta neurons.
Left: E-E coupling, Right: I-E coupling.\\

{\bf Figure 2.} (color online) (a) Activity diagrams for EE coupling:
Effect of different parameters on firing: (left: changing input $\beta_1=\beta_2=\beta$,
middle: changing coupling strength $g_{12}=g_{21}=g$, right: changing synaptic decay time
$\tau_{12}=\tau_{21}=\tau$.\\ 
(b) Plot in Fig.(2a, left) depicted in a more standard way, as variation of neuron 
outputs (on the y-axes) with the input $\beta$ (on the x-axis) showing bifurcation in 
the neuron activity.\\
 
{\bf Figure 3.}(color online) Complete synchronization in EE system in the presence of noise:
left: $\sigma=0.375$, ~right: $\sigma=1.0$.\\

{\bf Figure 4.}(color online) Noise produces delay in the decay time of the synaptic 
gating variables which eventually bring about CS in neurons with EE coupling. 
These plots illustrate this. 
(a) plots at the left are time-series of 
the synaptic conductances $s_{12}$ (in red) \& $s_{21}$ (blue) and the corresponding 
differences of the neuronal outputs $(u_1-u_2)$ are shown at the right for EE coupling. 
These plots show the approach to CS as $\sigma$ is increased from top to bottom: 
$\sigma=0$, $\sigma=0.20$, $\sigma=0.24$, $\sigma=0.26$, $\sigma=0.29002$, 
$\sigma=0.2900200338$.\\
(b) Detail of the plots on the left in (a) above are shown for increasing noise strengths 
($\sigma=0.0, 0.20, 0.24, 0.26$) for a single synaptic gating variable $s_{12}$.\\ 

{\bf Figure 5}(color online) Instantaneous frequency vs instantaneous phase of 
inputs to neurons (``flame plots'') for EE coupling for: 
(a) $\sigma=0.0$; (b) $\sigma=0.75$.\\
Parameters used are:
$\beta_1=0.0$,$\beta_2=0.00001$,$g_{12}=g_{21}=0.3$,
$\tau_{12}=\tau{21}=2.0,\tau_R=0.1,\eta=5.0$,
(Initial conditions: $\theta_1= 0.0, \theta_2= 0.0, s_{12}= 0.0, s_{21}= 0.0$).
-- Noise-induced CS occurs.\\

{\bf Figure 6}(color online) Firing frequency and Lyapunov exponents versus input curves for
EE coupling, for $\tau_{12}=\tau_{21}=2.0, \eta=5.0, \tau_R=0.1$, $\sigma=0$ :
(a) $g_{12}=g_{21}=0.3$ ;
(b) $g_{12}=0.3, g_{21}=0.45$.\\

{\bf Figure 7} (color online) Frequency of firing \& effect of noise on it for 
EE coupling for: (a) $\sigma=0.025$ \& $\sigma=0.05$,
(b) curves of the left graph plotted alongside values for $\sigma=0$.\\ 

{\bf Figure 8} (color online)
Variation of $\langle |u_1 - u_2| \rangle$  with the coupling strength
$g_{12}=g_{21}=g$ for EE coupling.  
Parameter values are:
$\tau_{12}=\tau_{21}= 2.0, \beta_1=\beta_2=0.1, \tau_{R}=0.1$.
(Initial conditions: $\theta_1=0.0,\theta_2=0.01,s_{12}=0.0,s_{21}=0.0$).\\

{\bf Figure 9} (color online)
Variation of $\langle |u_1 - u_2| \rangle$  with the input $\beta$ for EE coupling:
Parameter values are:
$\tau_{12}=\tau_{21}=2.0, g_{12}=g_{21}=0.3, \tau_{R}=0.1$.
(Initial conditions: $\theta_1=0.0,\theta_2=0.01,s_{12}=0.0,s_{21}=0.0$).\\

{\bf Figure 10} (color online)
(a) Variation of $\langle |u_1 - u_2| \rangle$  with the noise-strength $\sigma$
for EE coupling. (b) critical noise strength as a function of $g$ (eqn.(12));
(c) $\langle |u_1 - u_2| \rangle_{max}$ versus $g$ (eqn.(13)).\\
Parameter values are:
$\beta_1=\beta_2= 0.1, \tau_{12}=\tau_{21}=2.0, \tau_{R}=0.1$.
(Initial conditions: $\theta_1=0.0,\theta_2=0.01,s_{12}=0.0,s_{21}=0.0$).\\

{\bf Figure 11.} (color online) (a) Activity diagrams for IE coupling:
Effect of different parameters on firing: (left: changing input $\beta_1=\beta_2=\beta$,
middle: changing coupling strength $g_{12}=g_{21}=g$, right: changing synaptic decay time
$\tau_{12}=\tau_{21}=\tau$.\\
(b) Plot in Fig.(11a, left) depicted in the standard way, as variation of neuron 
outputs (on the y-axes) with the input $\beta$ (on the x-axis) showing bifurcation in 
the neuron activity.\\  

{\bf Figure 12.} (color online) Noise-induced order in IE systems:
for (from left to right): $\sigma=0.0, \sigma=0.5, \sigma=1.0$.\\

{\bf Figure 13.} (color online) IE system:
(a) relaxation oscillations of inputs; (b) canard in the $I_1$ - $I_2$ plane;
(c) time series of neuronal outputs $u_i$.
Here, $g_{12}=g_{21}=0.3, \tau_{12}=\tau_{21}=2.0, \tau_R=0.1,
\eta=5.0, \beta_1=\beta_2=0.1, \alpha_1=+1, \alpha_2=-1$.\\

{\bf Figure 14}(color online) Instantaneous frequency vs instantaneous phase of inputs 
to neurons (``flame plots'') for IE coupling for:\\ 
(a) $\sigma=0.0$, $\beta_1=\beta_2=0.1$,
$g_{12}=g_{21}=0.3$,$\tau_{12}=\tau{21}=2.0,\tau_R=0.1,\eta=5.0$ 
(Initial conditions: $\theta_1=0.0,\theta_2=0.1,s_{12}=s_{21}=0.0$);
(b) $\sigma=0.15$, $\beta_1=\beta_2=0.0$,
$g_{12}=g_{21}=0.3$,$\tau_{12}=\tau{21}=2.0,\tau_R=0.1,\eta=5.0$, 
(Initial conditions: $\theta_1=0.0,\theta_2=0.01,s_{12}=0.0,s_{21}=0.0$).\\
CS does not occur for this case.\\

{\bf Figure 15}(color online) (a) Firing frequency and (b) Lyapunov exponents 
versus input curves for IE coupling for  
$\tau_{12}=\tau_{21}=2.0, \eta=5.0, \tau_R=0.1$, $\sigma=0$, $g_{12}=g_{21}=0.3$.\\

{\bf Figure 16} (color online) Frequency of firing as a function of constant input 
\& effect of noise on it for IE coupling : for $\sigma=0, \sigma=0.025, \sigma=0.075$.
(a) inhibitory neuron, (b) excitatory neuron).\\

{\bf Figure 17} (color online)
Variation of $\langle |u_1 - u_2| \rangle$  with the coupling strength
$g_{12}=g_{21}=g$ for IE coupling.  
Parameter values are:
$\tau_{12}=\tau_{21}= 2.0, \beta_1=\beta_2=0.1, \tau_{R}=0.1$.
(Initial conditions: $\theta_1=0.0,\theta_2=0.01,s_{12}=0.0,s_{21}=0.0$).\\

{\bf Figure 18} (color online)
Variation of $\langle |u_1 - u_2| \rangle$  with the input $\beta_1=\beta_2=\beta$
for IE coupling. (a) The curves correspond to eqn.(16) ($\beta>\beta_c$):
$k, l$ values associated with
a given $\sigma$ are: Black: $\sigma=0.0, k=0.0564, l=-1.6731$,
Red: $\sigma=0.3, k=0.0281, l=-1.3538$,  Green: $\sigma=0.6, k=0.0092, l=-1.0569$,
Blue: $\sigma=1.0, k=-0.0052, l=-0.6903$ ; (b) Same as (a), but on a log-log plot, 
clearly showing cross-over behaviour. (c)Detail of plot on the left for
$\beta \le \beta_c$ ,curves correspond to eqn.(15) : $a, b$ values associated
with a given $\sigma$ are: Black: $\sigma=0.0, a=0.9820, b=0.5042$,
Red: $\sigma=0.3, a=0.3305, b=0.1913$, Green: $\sigma=0.6, a=0.2150, b=0.0481$,
Blue: $\sigma=1.0, a=0.2215, b=0.0460$. \\
Parameter values are:
$\tau_{12}=\tau_{21}=2.0, g_{12}=g_{21}=0.25, \tau_{R}=0.1$.
(Initial conditions: $\theta_1=0.0,\theta_2=0.01,s_{12}=0.0,s_{21}=0.0$).\\

{\bf Figure 19} (color online)
Variation of $\langle |u_1 - u_2| \rangle$  with the noise-strength $\sigma$
for IE coupling.
Parameter values are:
$\beta_1=\beta_2=0.1, \tau_{12}=\tau_{21}=2.0, \tau_{R}=0.1$.
(Initial conditions: $\theta_1=0.0,\theta_2=0.01,s_{12}=0.0,s_{21}=0.0$).\\

{\bf Figure 20} (color online) Rate of complete synchronization $T_{sync}$ with noise
keeping all other parameters fixed, for the EE system.

%\newpage
\begin{center}
\includegraphics [height=4.5cm,width=8.5cm,angle=0]{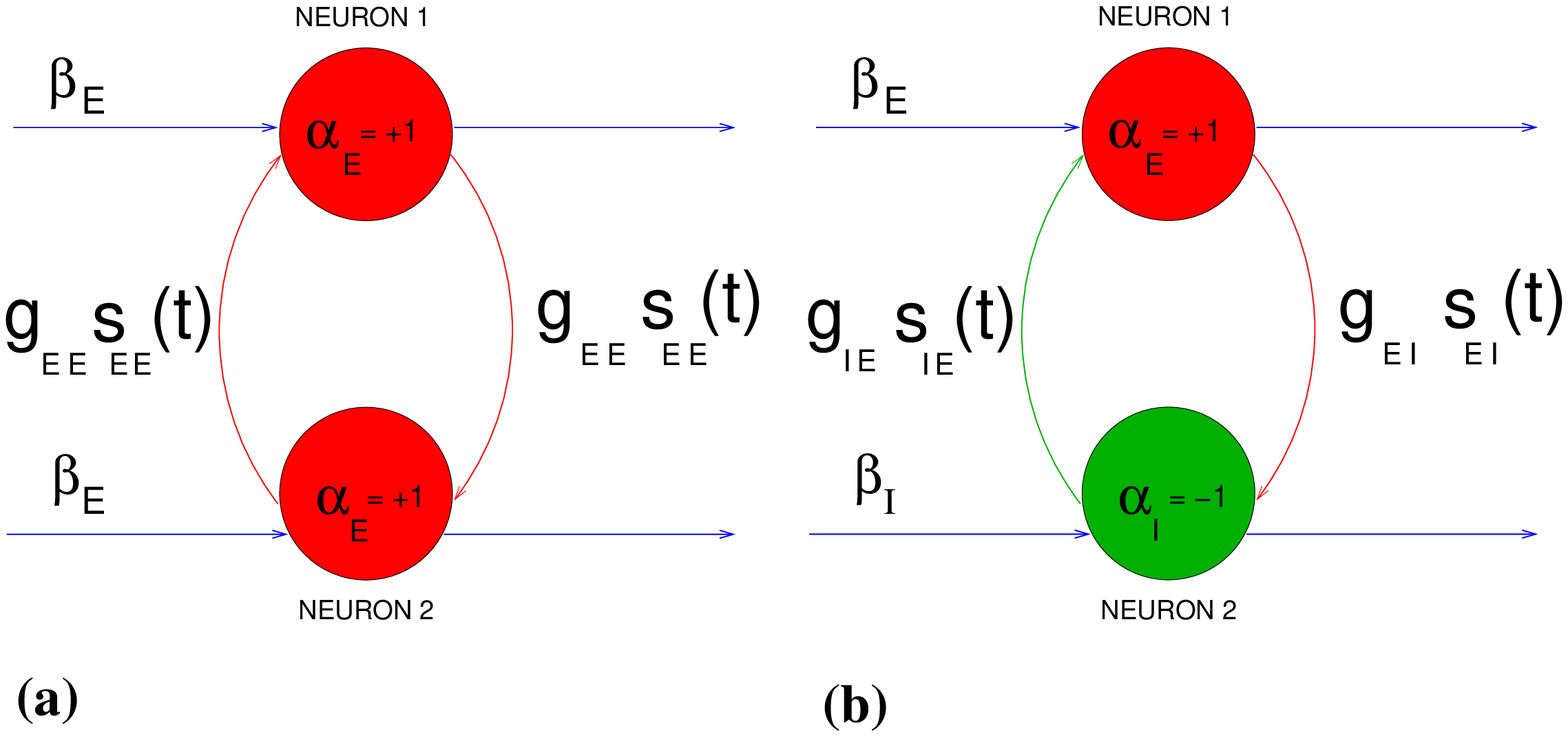}\\
\end{center}
{\bf Fig. 1} (color online)
%\newpage
\begin{center}
\includegraphics[width=8cm,height=5cm]{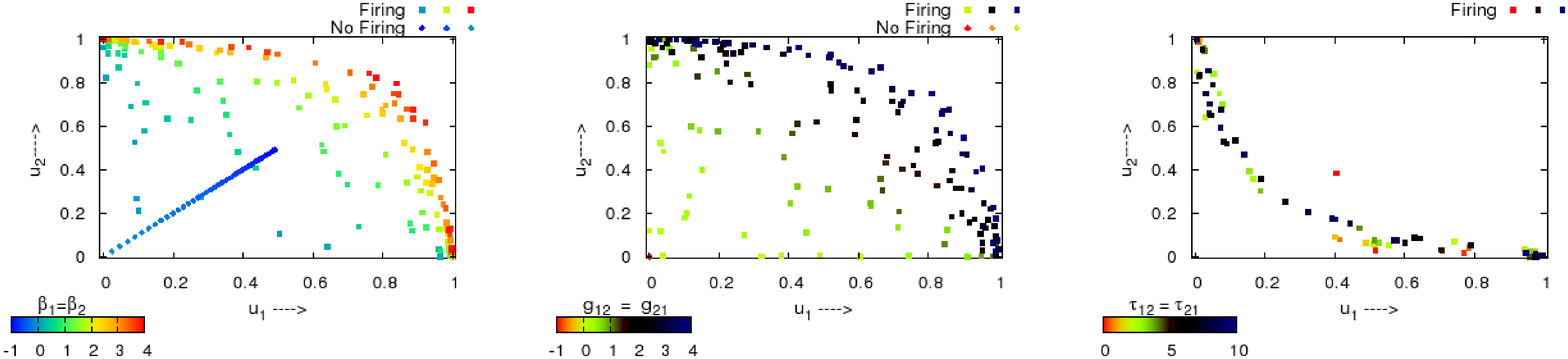}\\
\end{center}
{\bf Fig. 2a} (color online)

\begin{center}
\includegraphics[width=6cm,height=8.2cm,angle=270]{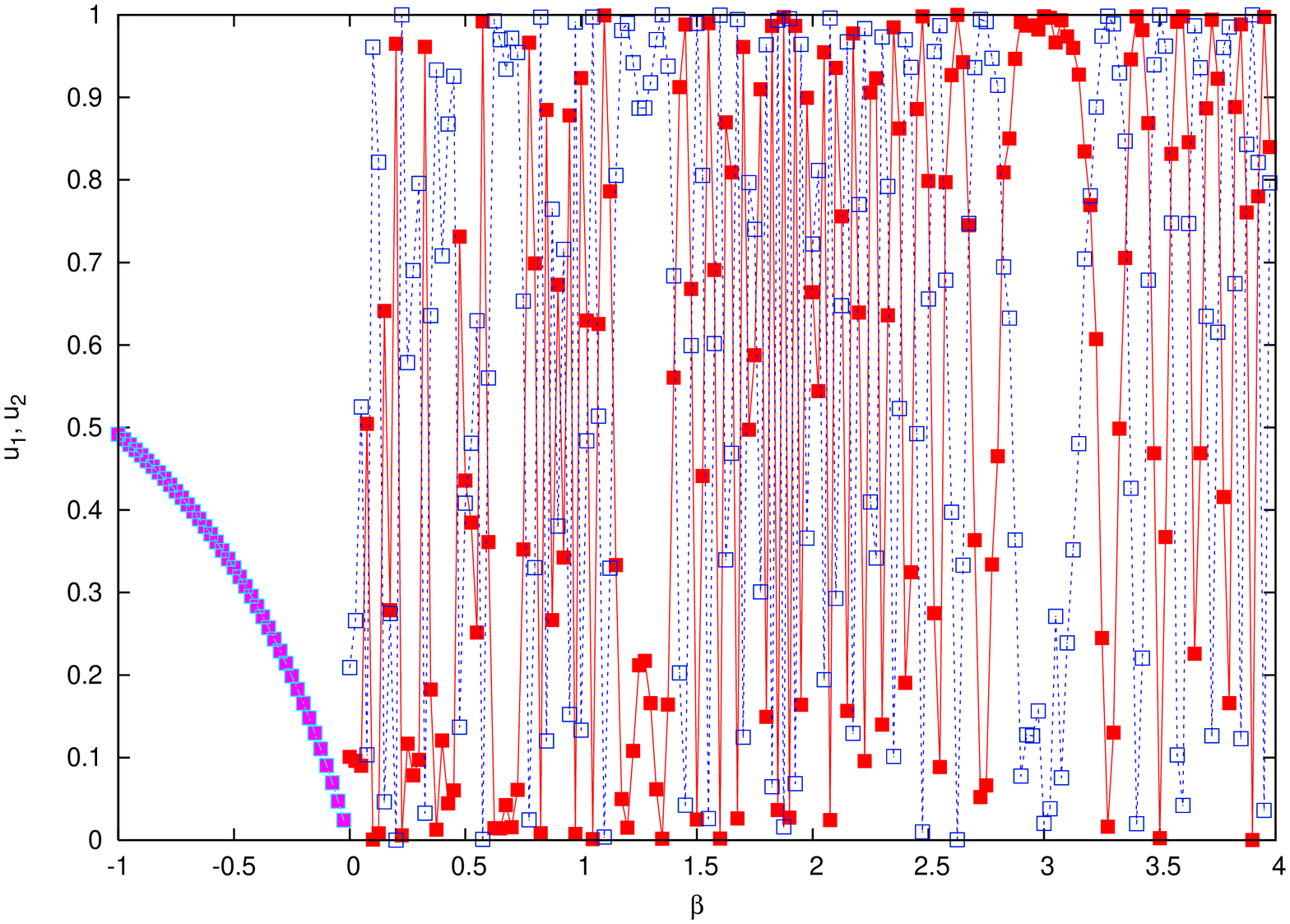}\\ 
\end{center}
{\bf Fig. 2b} (color online)
%\newpage
\begin{center}
\includegraphics[width=8cm,height=5.5cm,angle=0]{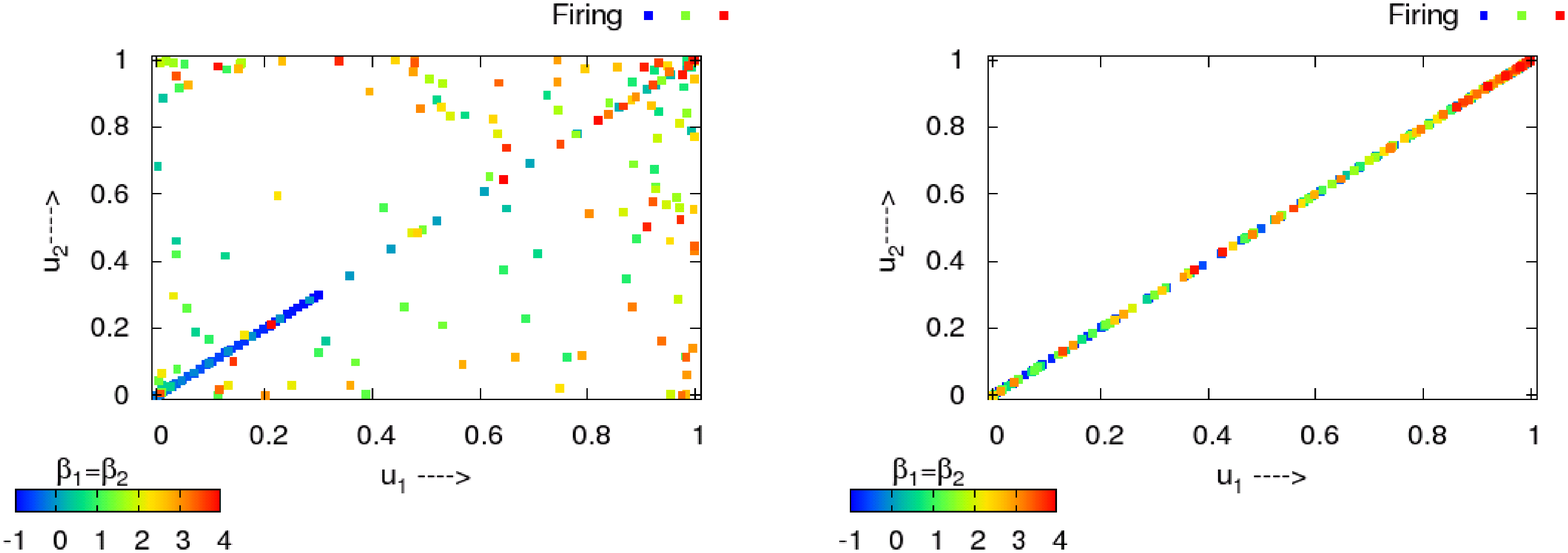}\\
\end{center}
{\bf Fig. 3} (color online)
\newpage
\begin{center}
\includegraphics[width=16cm,height=23cm,angle=0]{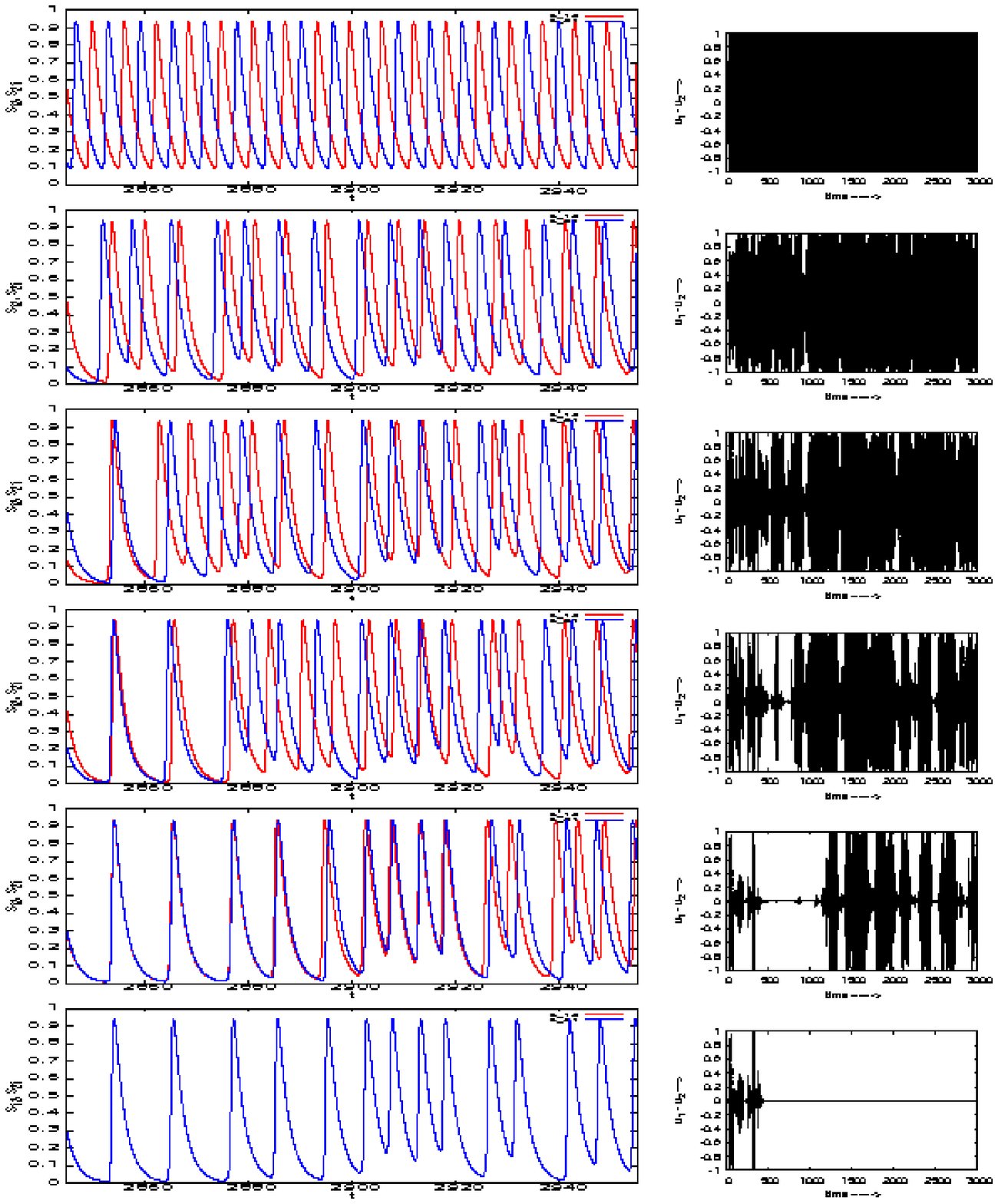}\\
\end{center}
{\bf Fig. 4a} (color online)
\newpage
\begin{center} 
\includegraphics[width=8cm,height=5.5cm,angle=270]{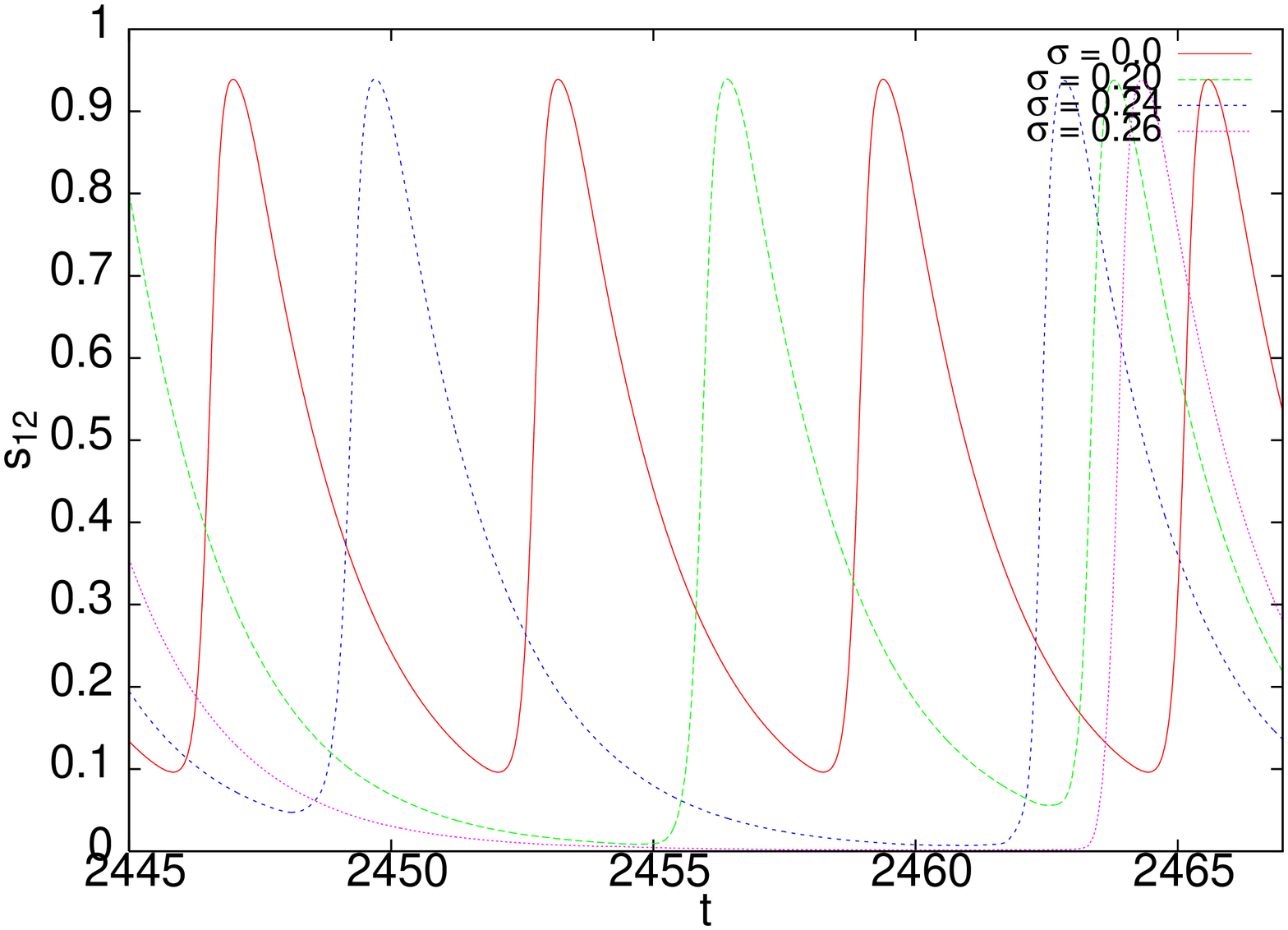}\\
\end{center}
{\bf Fig. 4b} (color online)
%\newpage
\begin{center}
\includegraphics[width=8cm,height=7cm,angle=0]{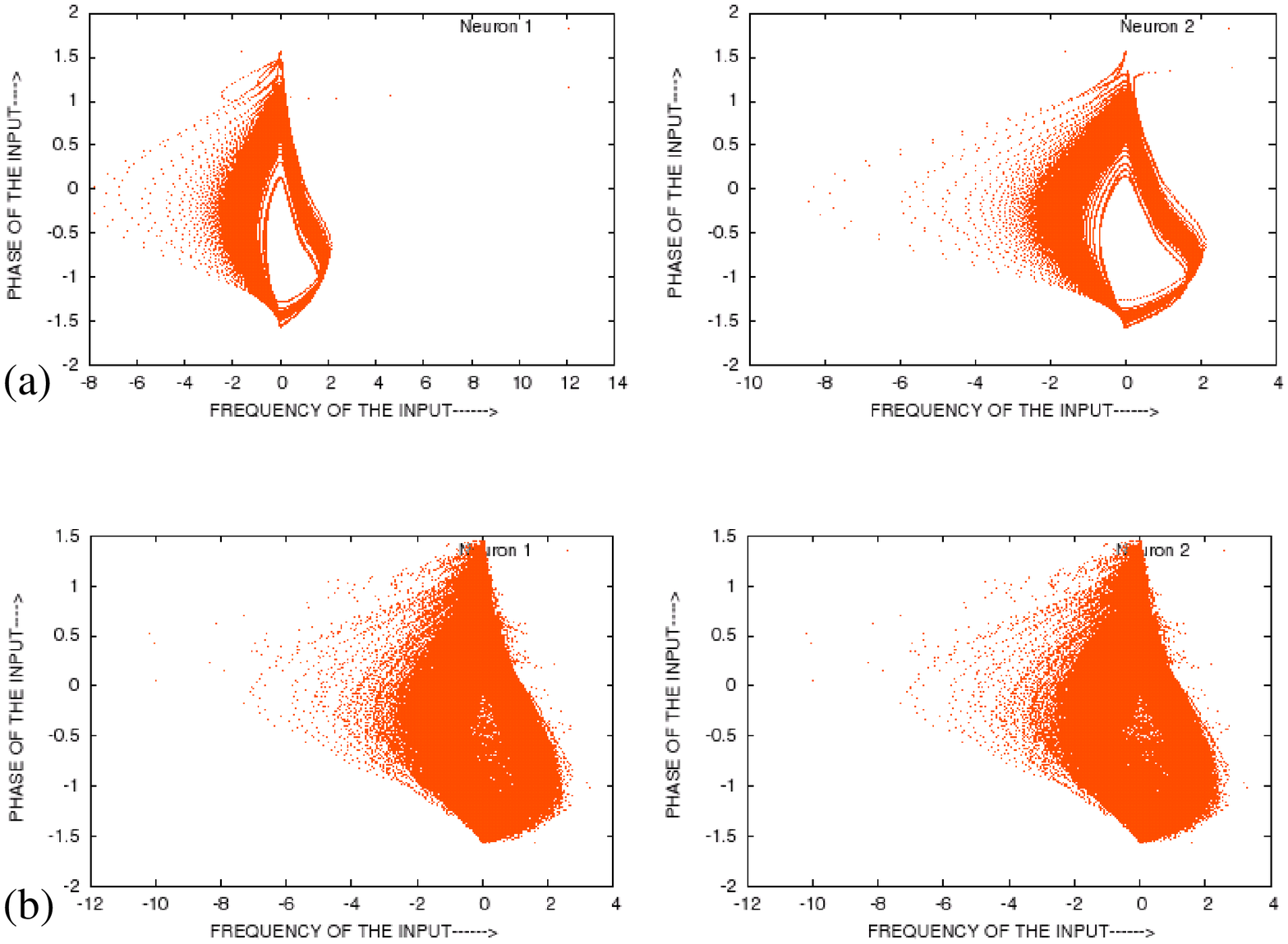}\\
\end{center}
{\bf Fig. 5} (color online)
%\newpage
\begin{center}
\includegraphics[width=8cm,height=7cm,angle=0]{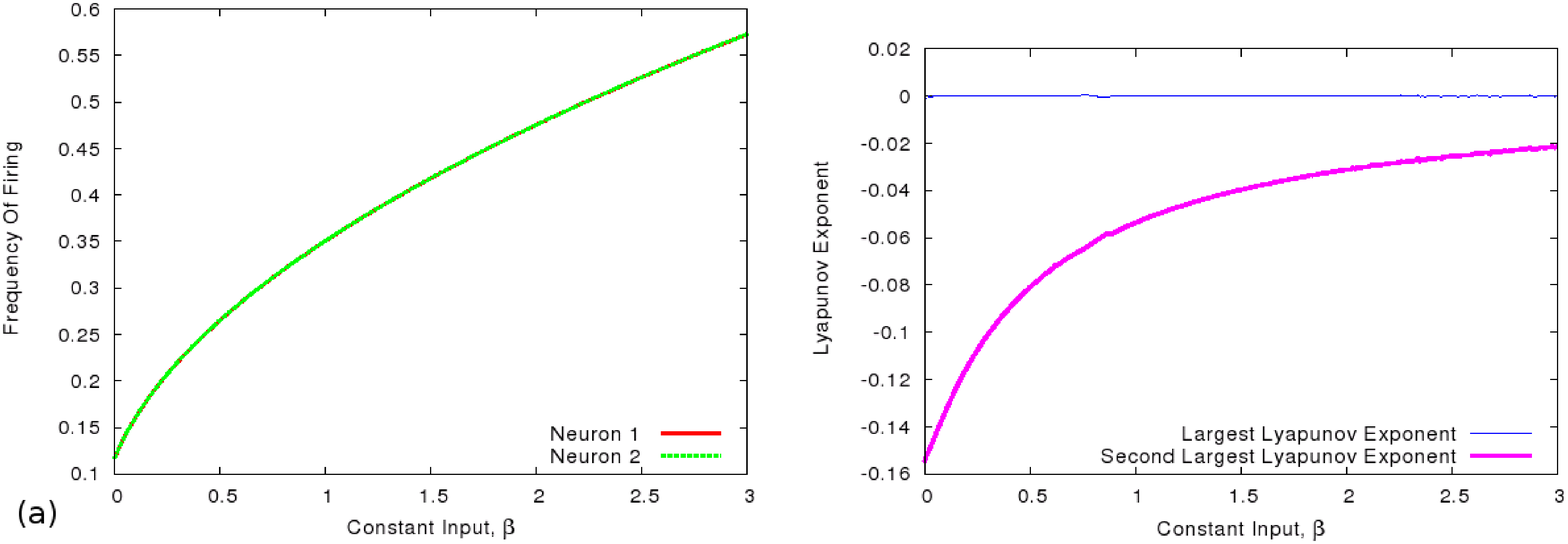}\\
\includegraphics[width=8cm,height=7cm,angle=0]{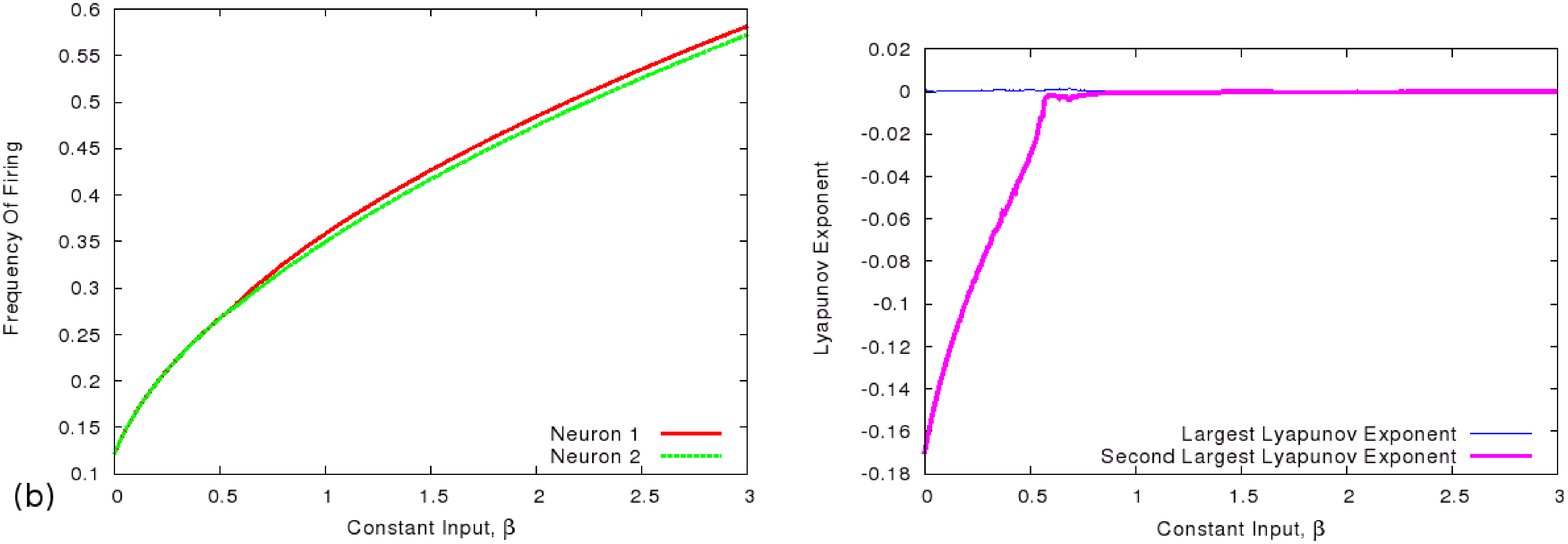}\\
\end{center}
{\bf Fig. 6} (color online)
%\newpage
\begin{center}
\includegraphics[width=8cm,height=5cm,angle=0]{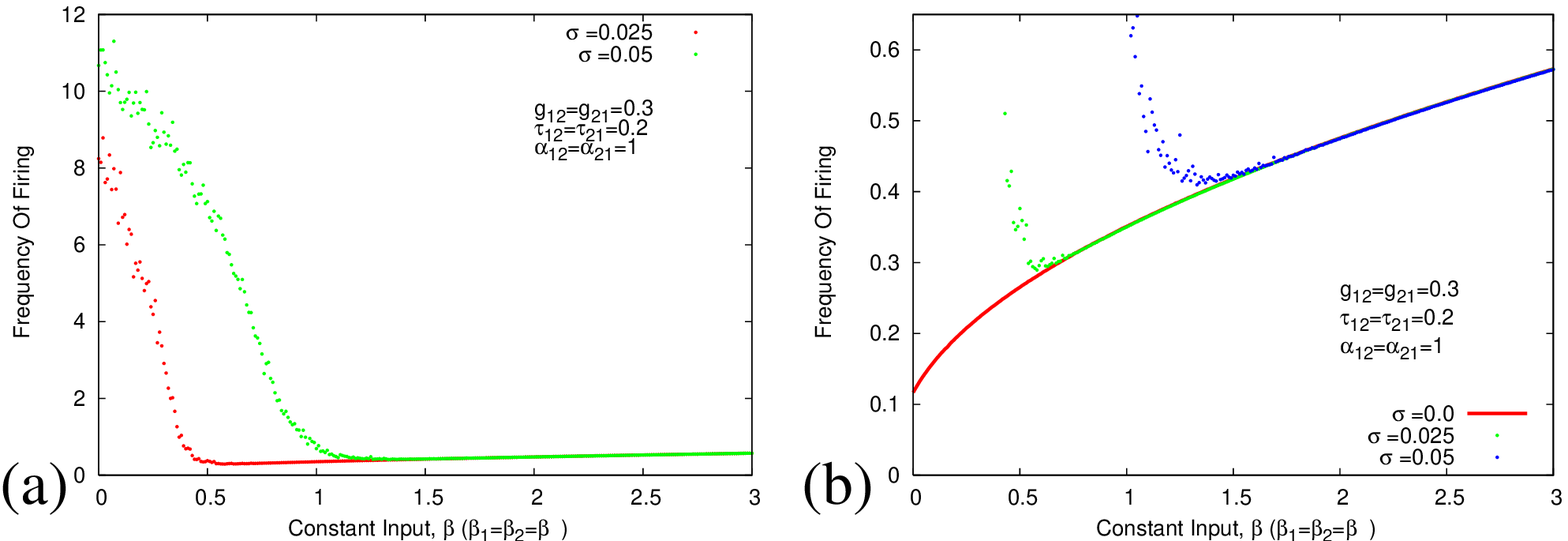}\\
\end{center}
{\bf Fig. 7} (color online)
\newpage
\begin{center}
\includegraphics[width=5cm,height=6cm,angle=270]{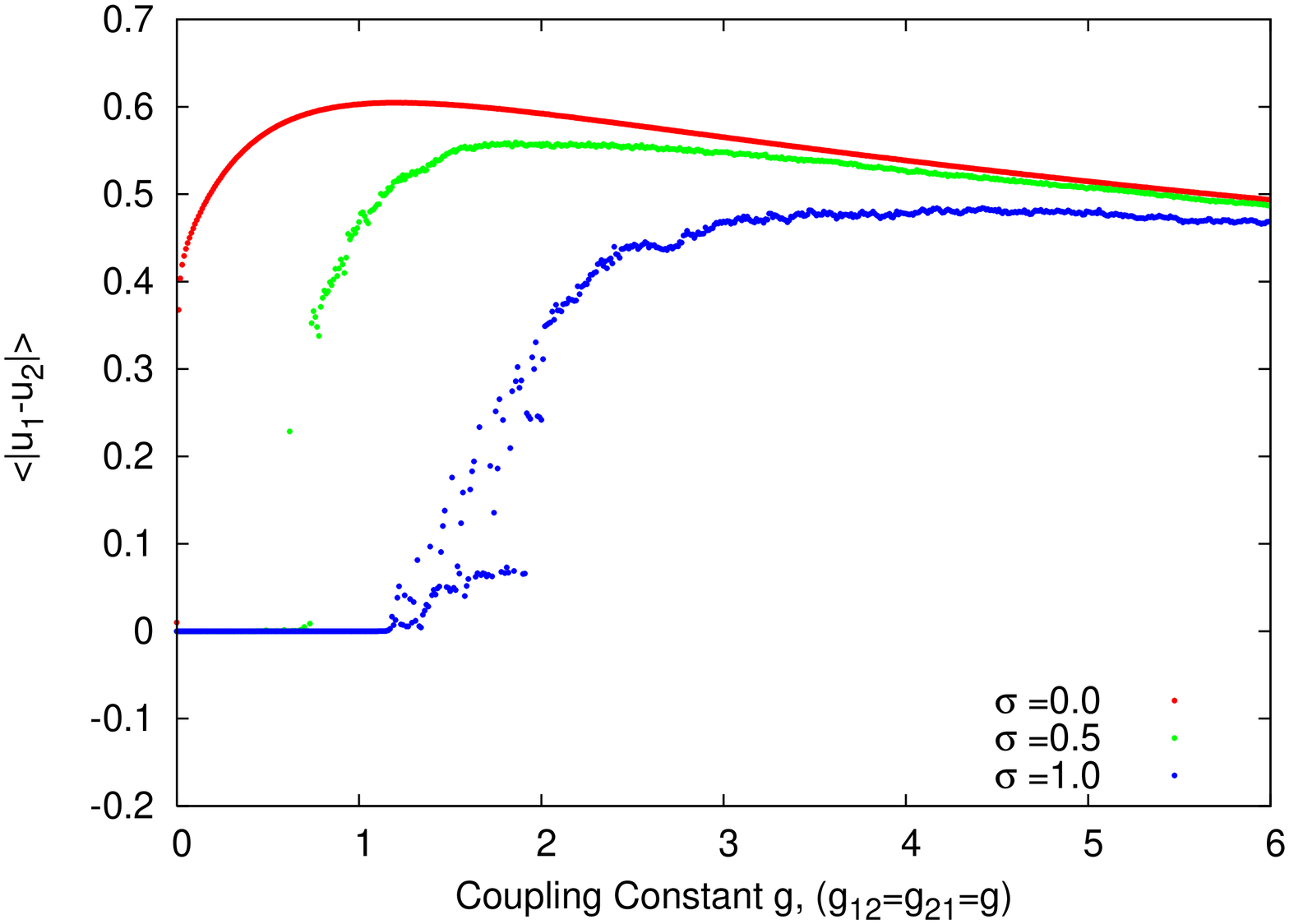}\\
\end{center}
{\bf Fig. 8}(color online)
%\newpage
\begin{center}
\includegraphics[width=5cm,height=6cm,angle=270]{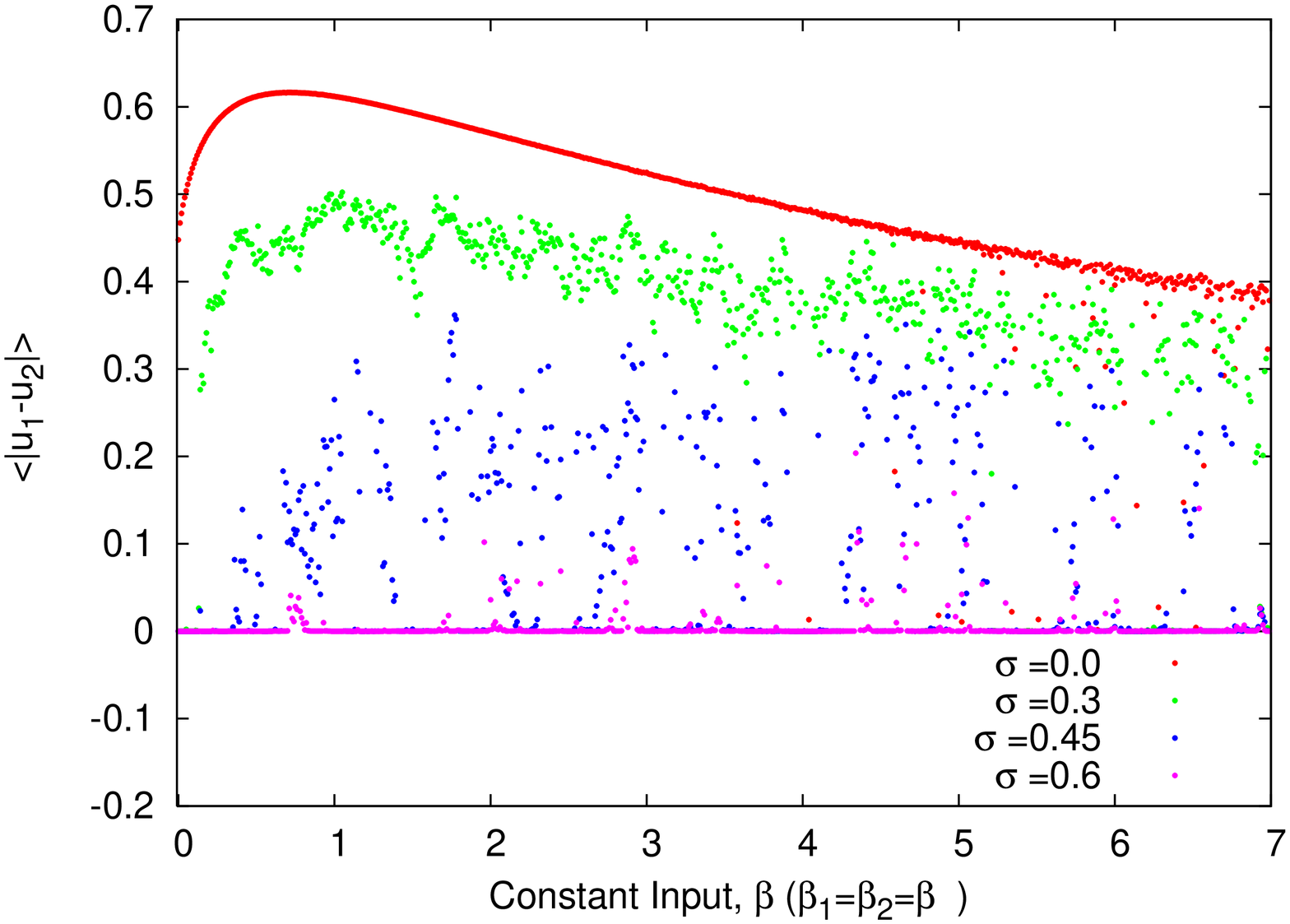}\\
\end{center}
{\bf Fig. 9}(color online)
\newpage 
\begin{center}
{\bf (a)}\hspace*{0.1cm}
\includegraphics[width=8.5cm,height=4.5cm,angle=0]{fig.10a.eps}\\
{\bf (b)}\hspace*{0.1cm}
\includegraphics[width=8.5cm,height=4.5cm,angle=0]{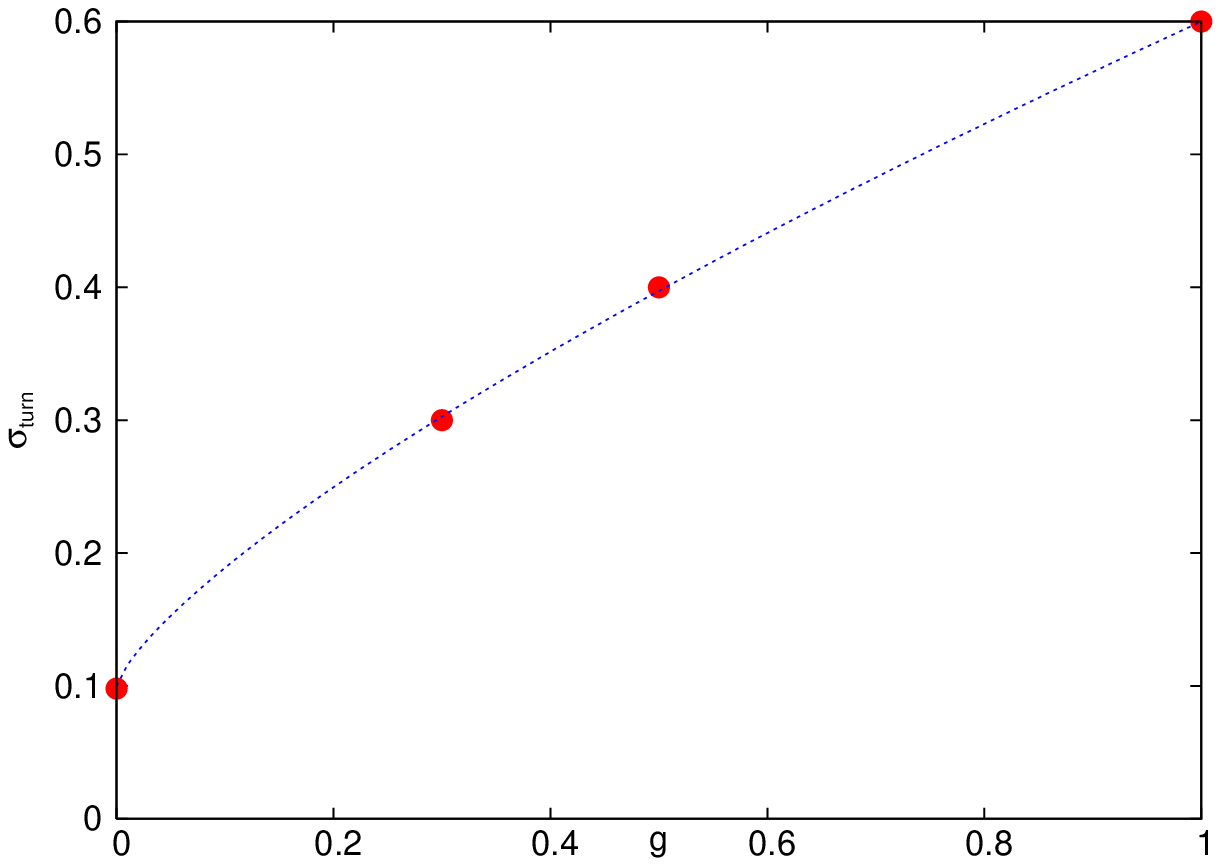}\\
{\bf (c)}\hspace*{0.1cm}
\includegraphics[width=8.5cm,height=4.5cm,angle=0]{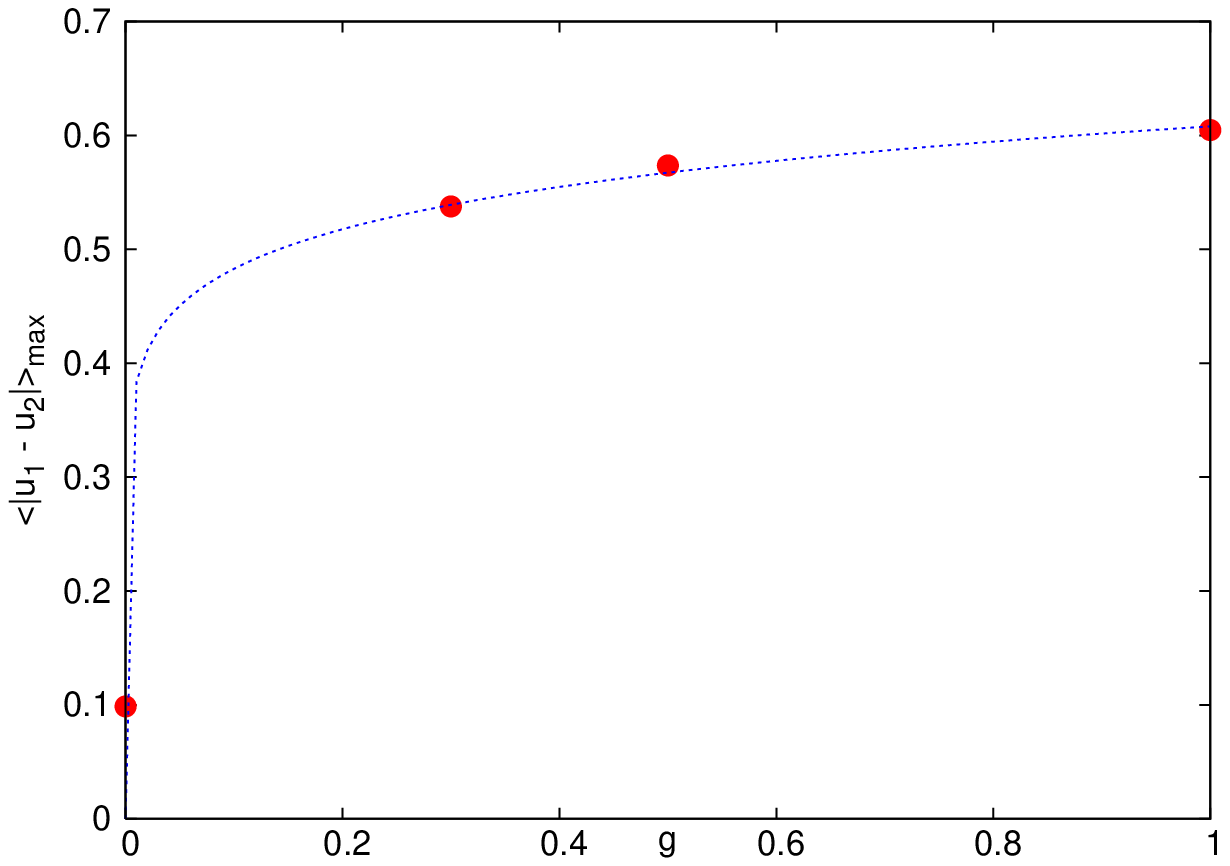}\\
\end{center}
\vspace*{1cm}
{\bf Fig. 10} (color online)
\newpage
\begin{center}
\includegraphics[width=8.5cm,height=5cm,angle=0]{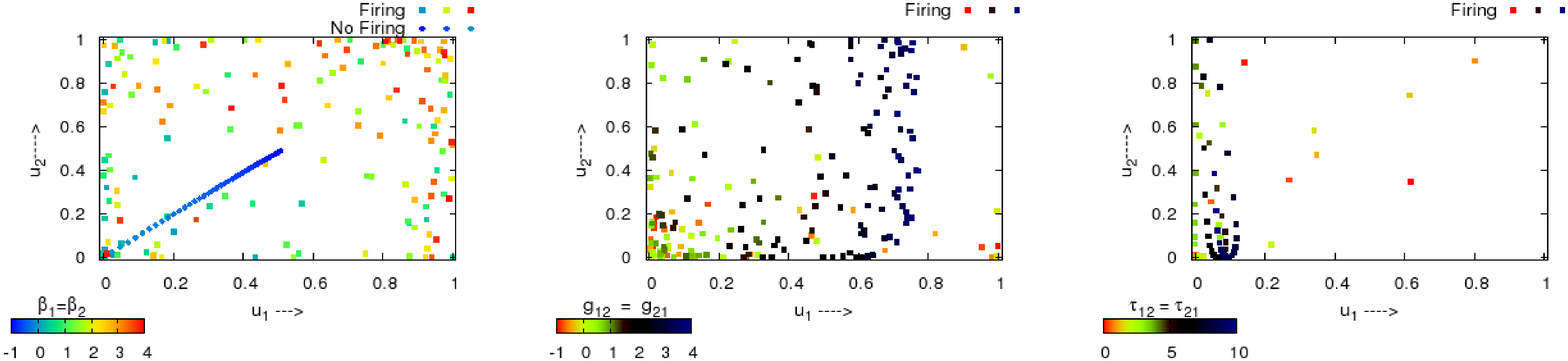}\\
\end{center}
{\bf Fig. 11a} (color online)
\begin{center} 
\includegraphics[width=6cm,height=8.2cm,angle=270]{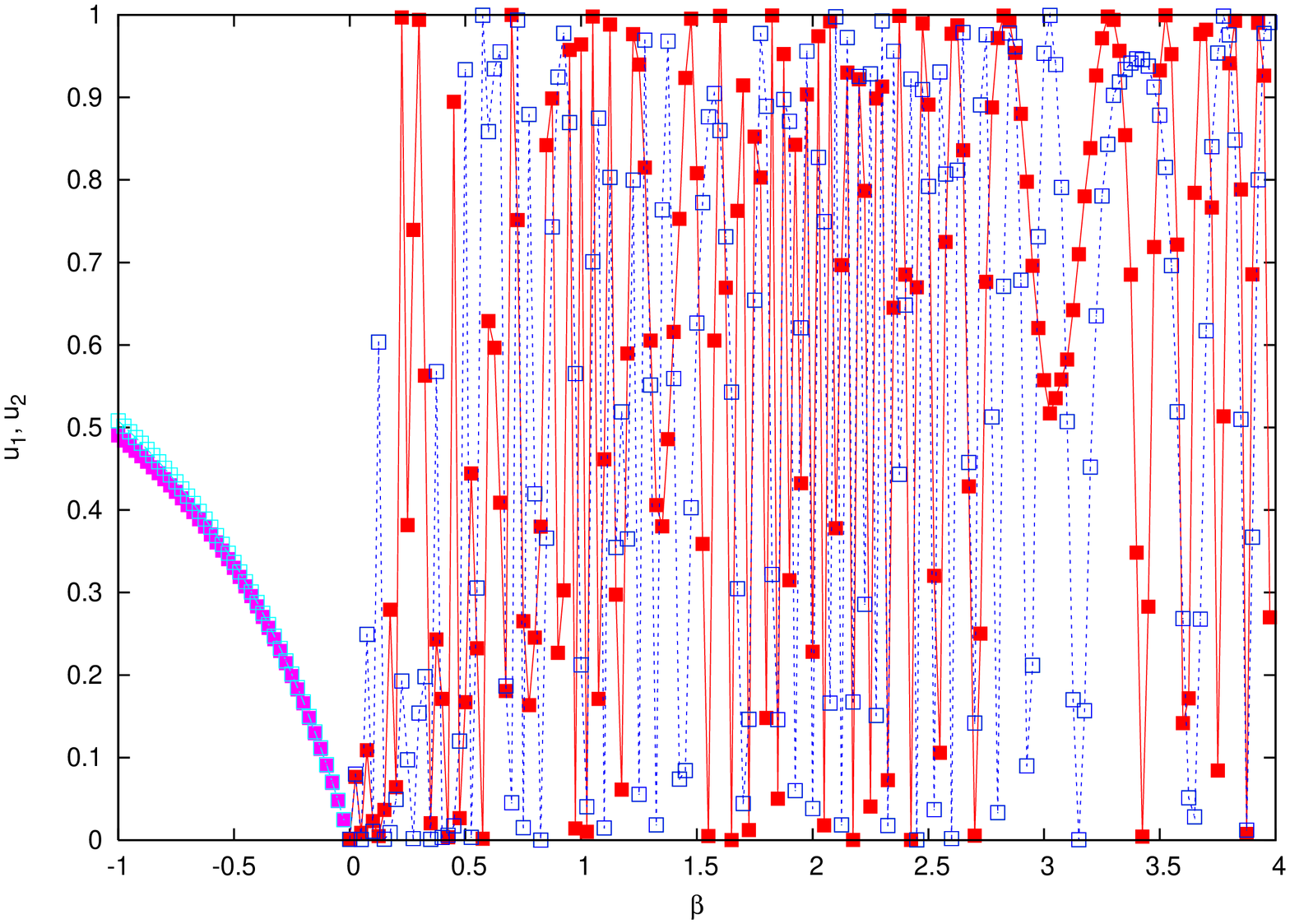}\\
\end{center}
{\bf Fig. 11b} (color online)
%\newpage
\begin{center}
\includegraphics[width=8.5cm,height=5cm,angle=0]{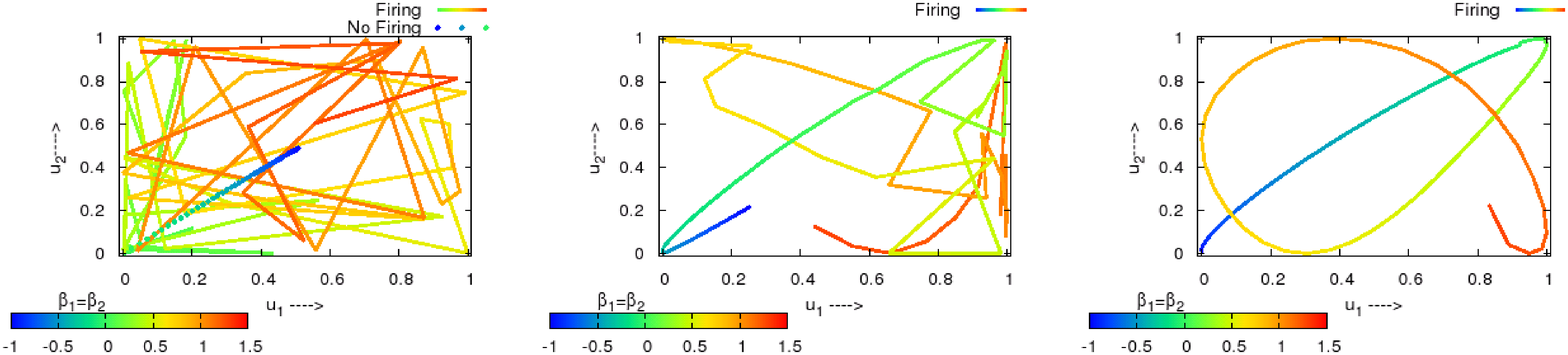}\\
\end{center}
{\bf Fig. 12} (color online)
\newpage
\begin{center}
\includegraphics[width=8.5cm,height=4.5cm,angle=0]{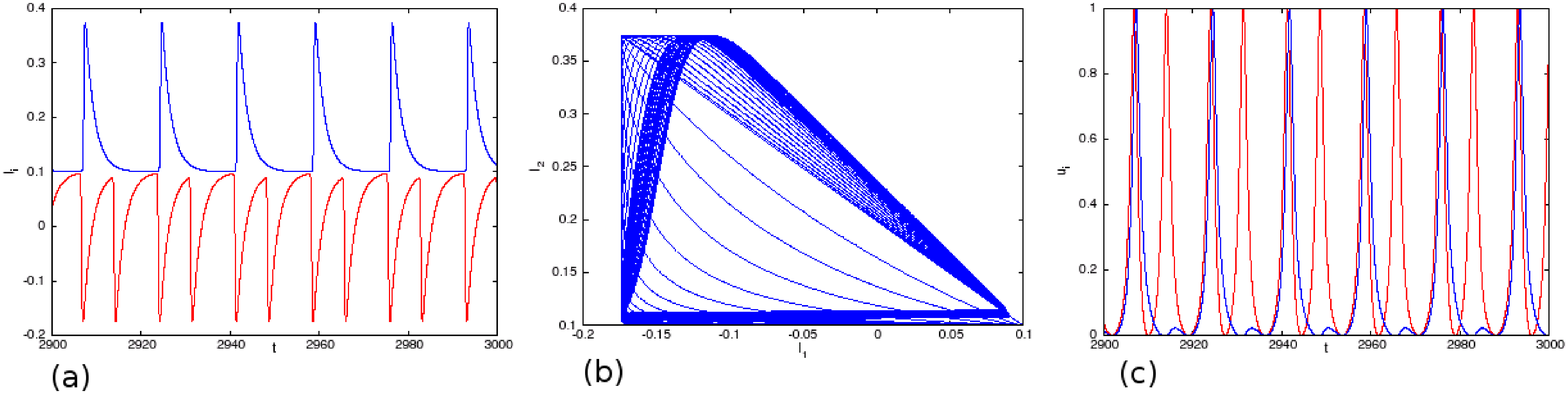}\\
\vspace*{1cm}
\end{center}
{\bf Fig. 13} (color online)
%\newpage
\begin{center}
\includegraphics[width=8cm,height=4.5cm,angle=0]{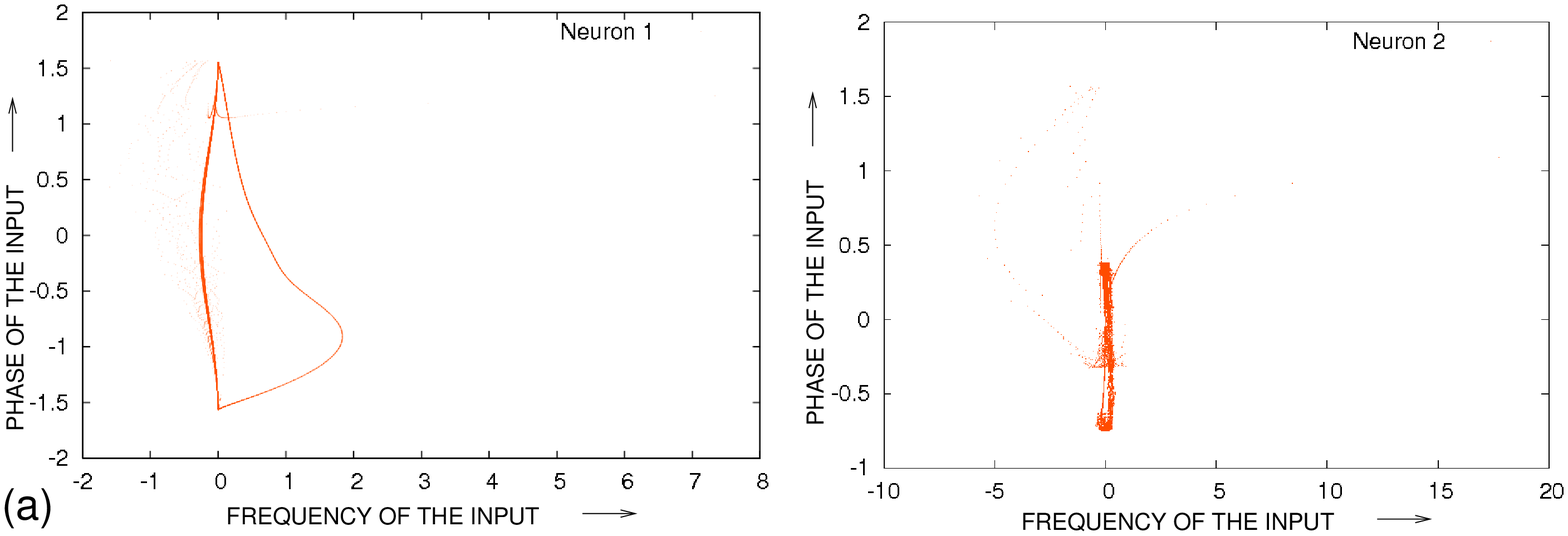}\\
\includegraphics[width=8cm,height=4.5cm,angle=0]{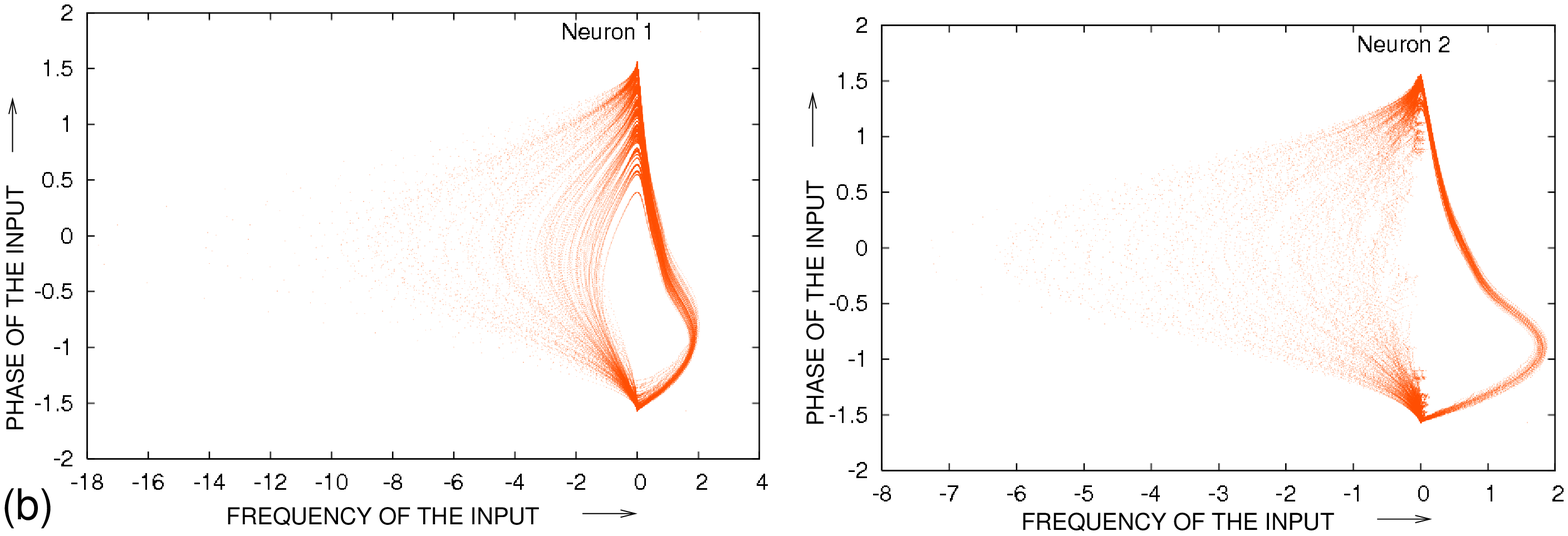}\\ 
\end{center}
{\bf Fig. 14} (color online)
\newpage
\begin{center} 
\includegraphics[width=8.5cm,height=5cm,angle=0]{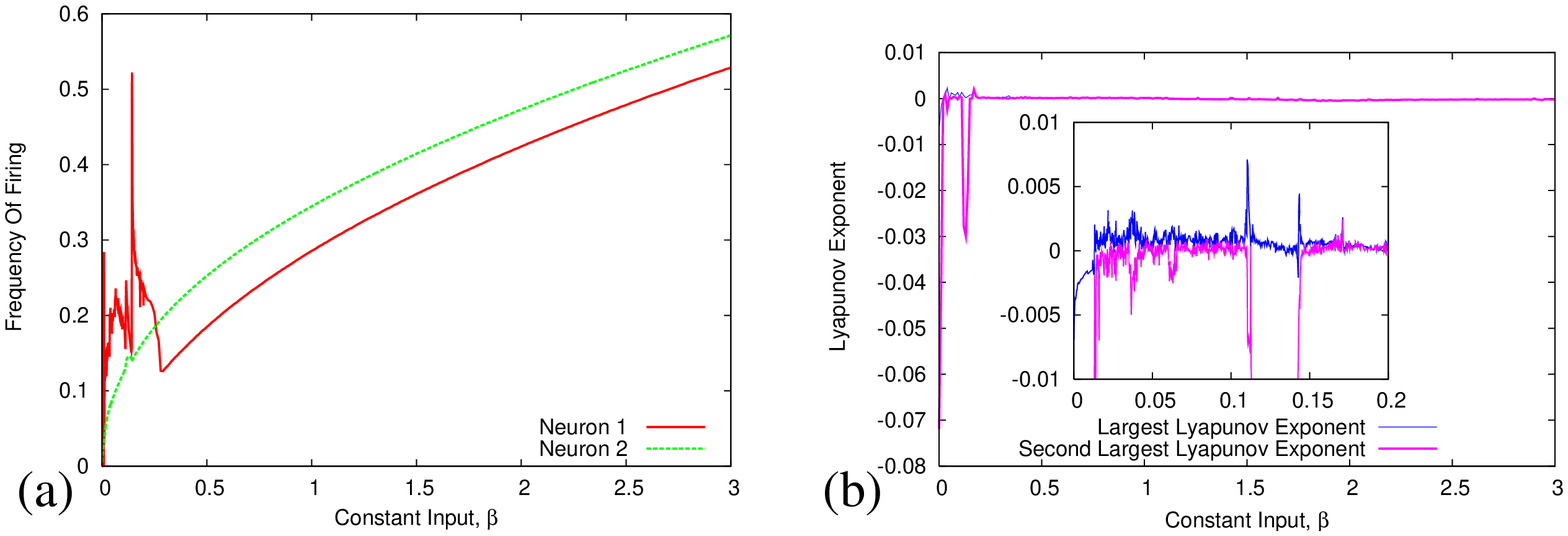}\\
\end{center}
{\bf Fig. 15} (color online)
%\newpage
\begin{center}
\includegraphics[width=8.5cm,height=5cm,angle=0]{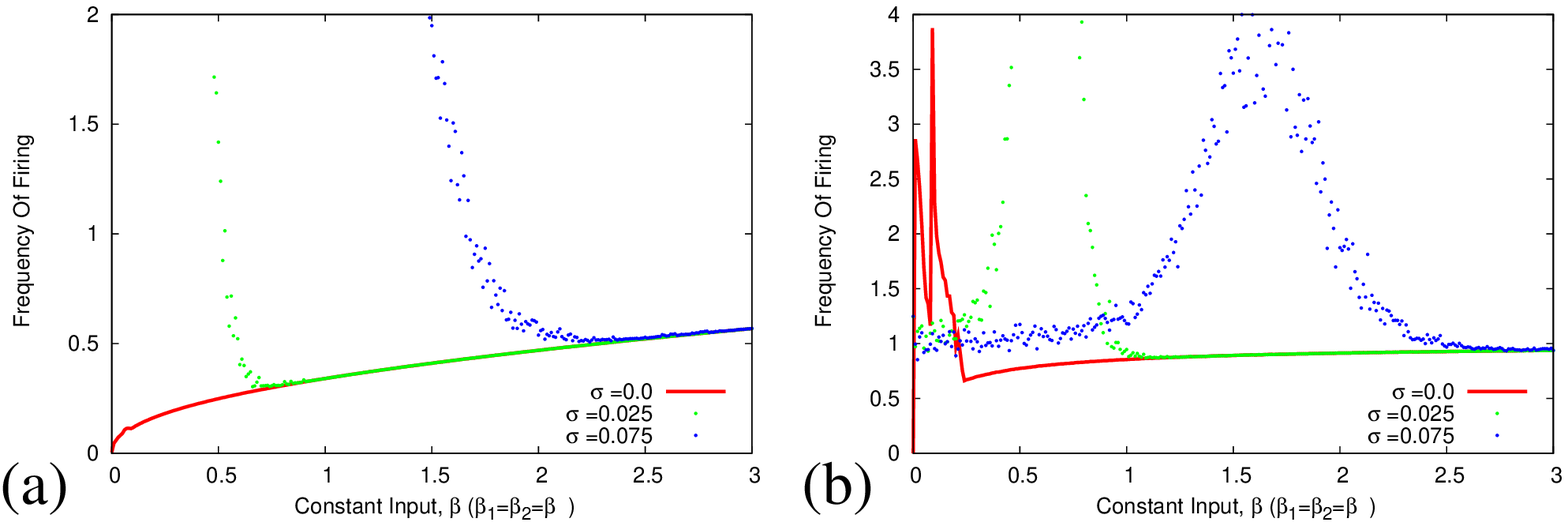}\\
\end{center}
{\bf Fig. 16} (color online)
%\newpage
\begin{center} 
\includegraphics[width=5cm,height=8.5cm,angle=270]{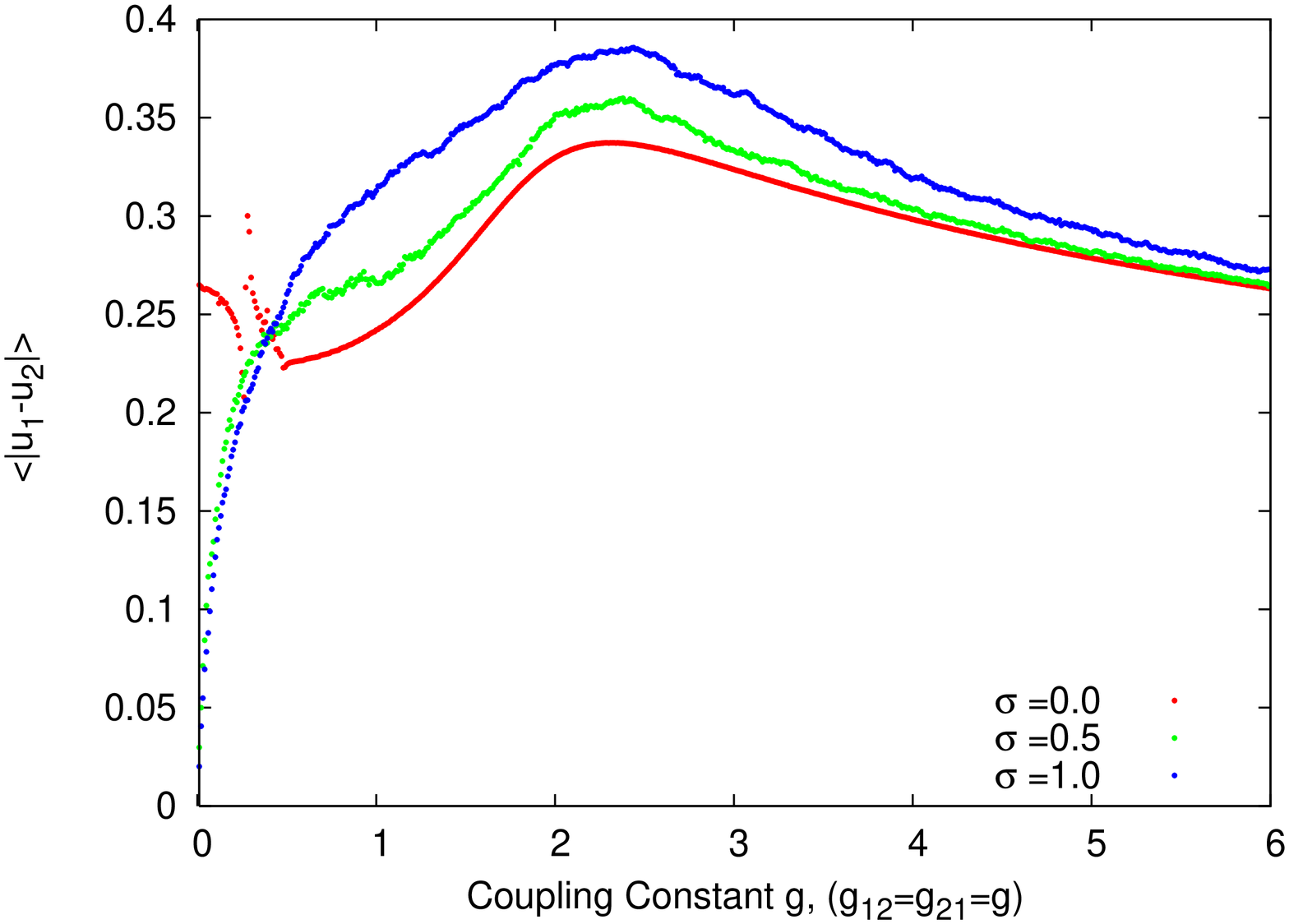}\\
\end{center}
{\bf Fig. 17}(color online)
\newpage 
\begin{center} 
\hspace*{0.1cm}
\includegraphics[width=9.2cm,height=7cm,angle=0]{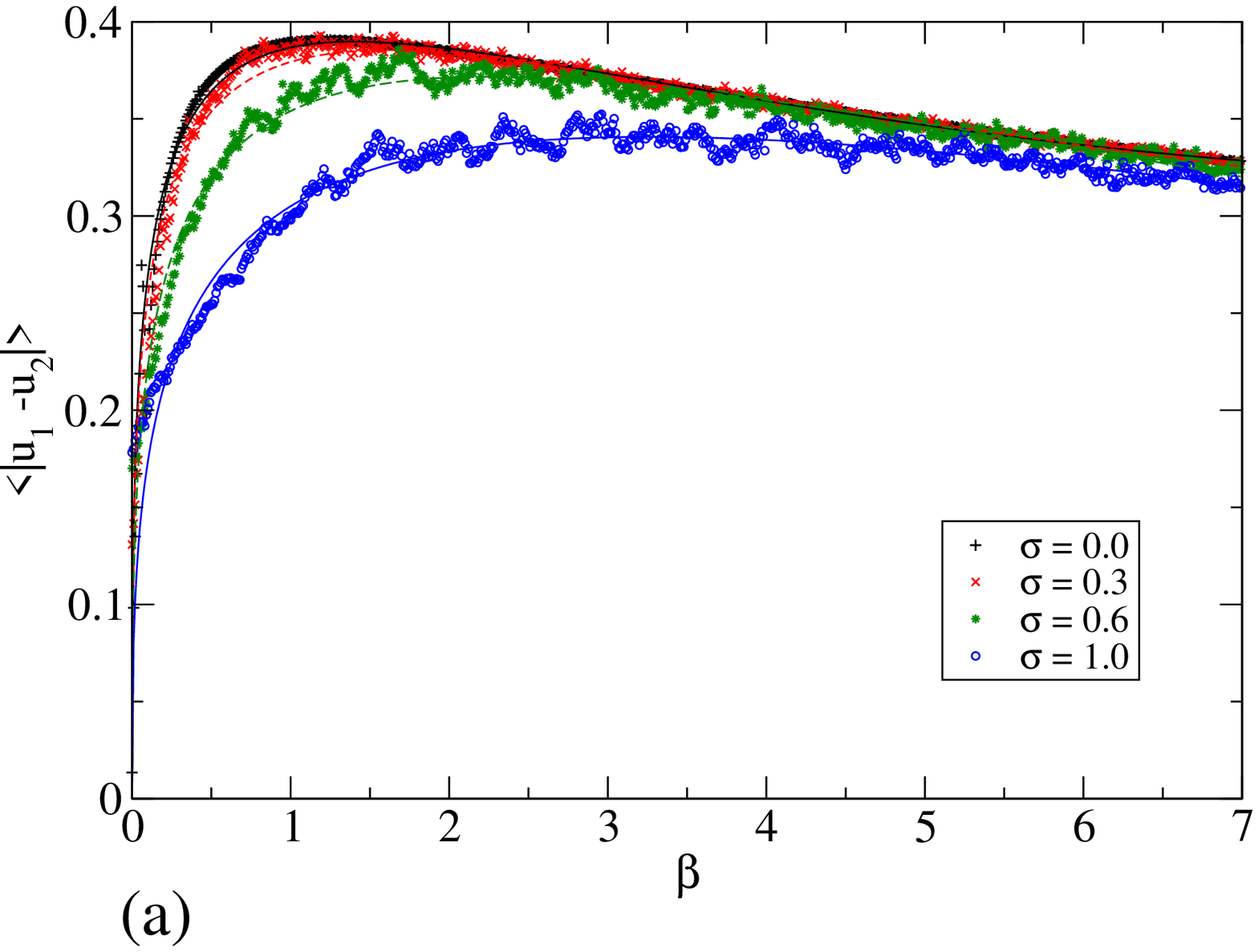}\\
\hspace*{0.1cm}
\includegraphics[width=9.2cm,height=7cm,angle=0]{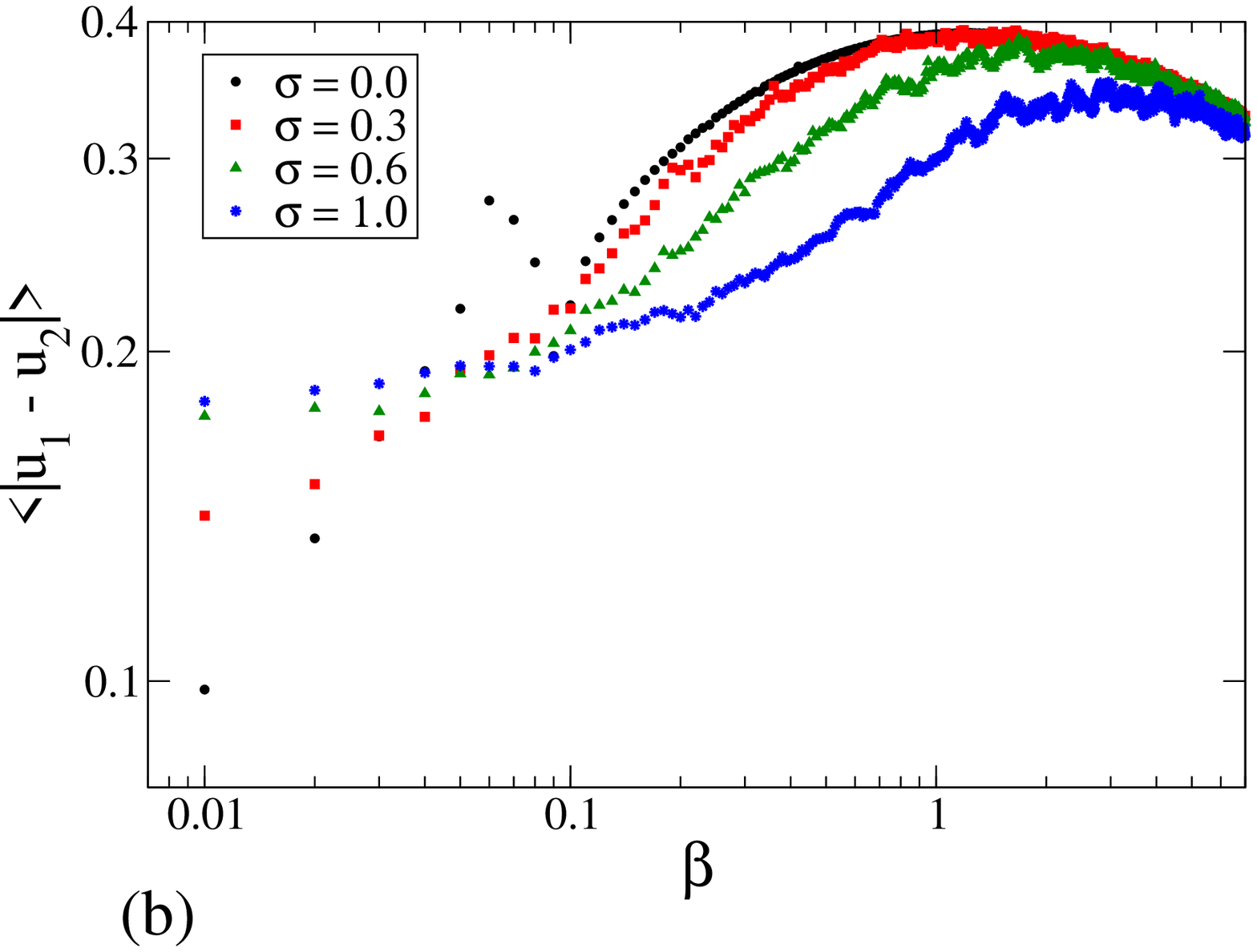}\\
\hspace*{0.1cm}
\includegraphics[width=9.2cm,height=7cm,angle=0]{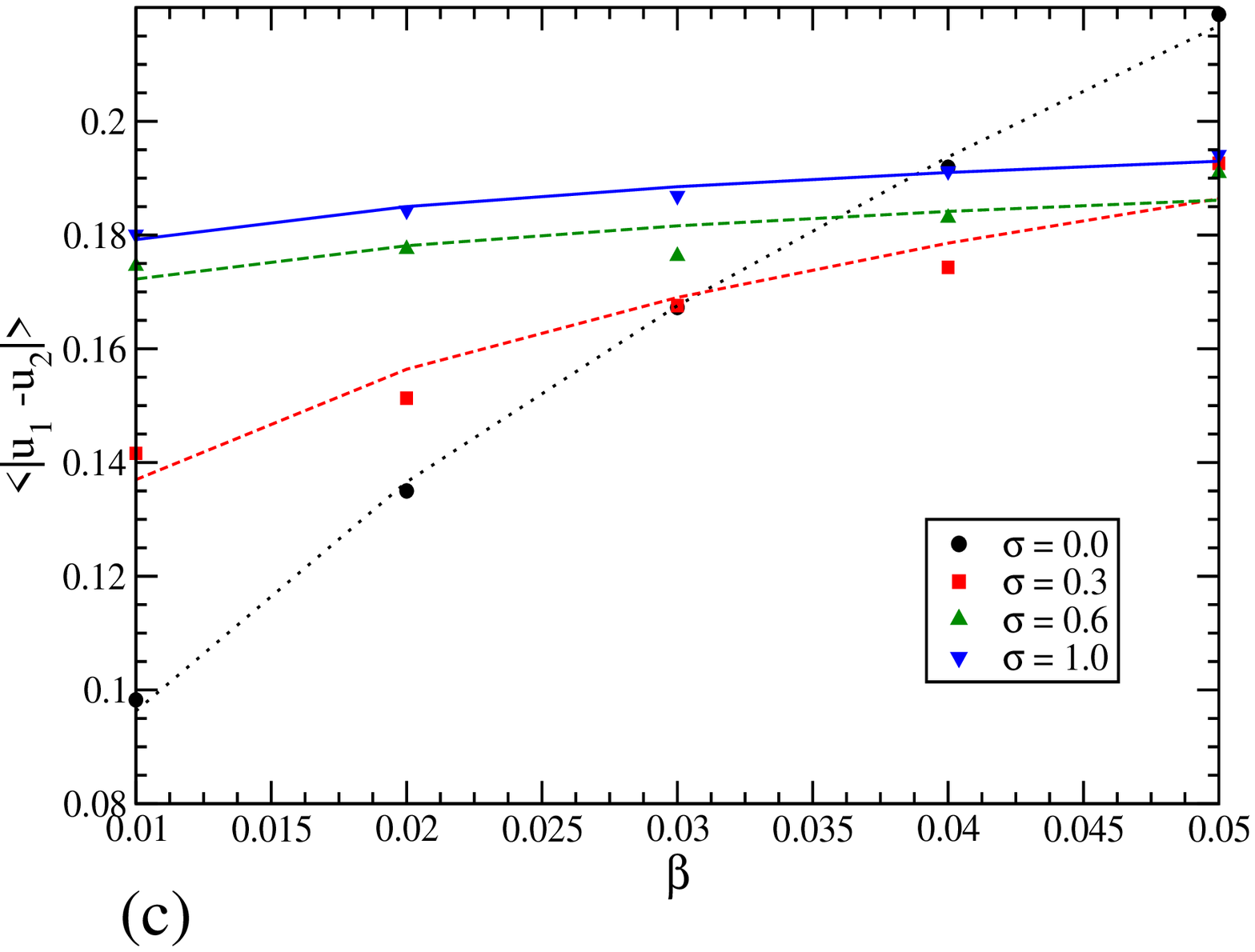}\\
\end{center}
\vspace*{1cm}
{\bf Fig. 18} (color online) 
\newpage
\begin{center}
\includegraphics[width=6cm,height=8.5cm,angle=270]{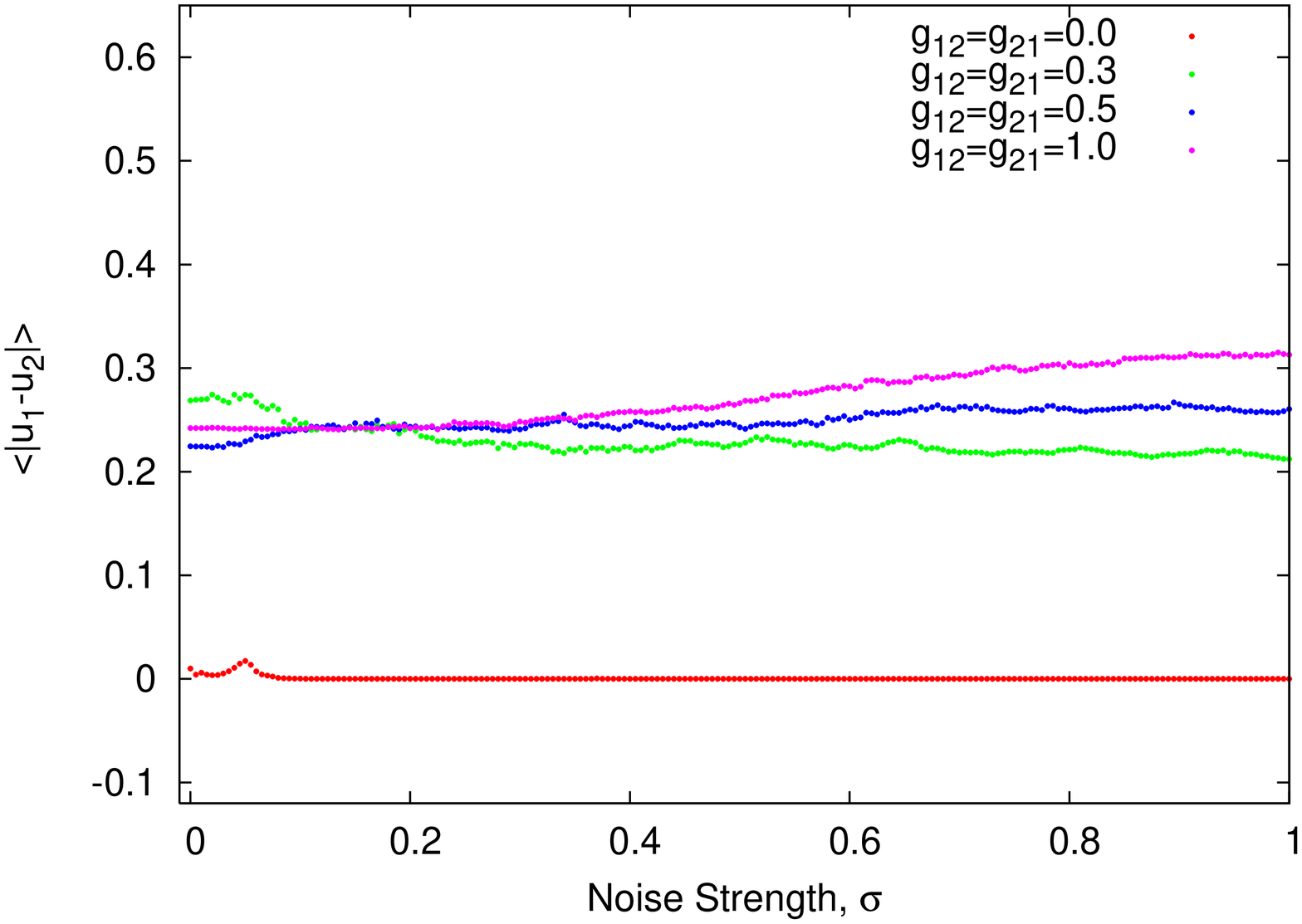}\\
\end{center}
{\bf Fig. 19} (color online)
\newpage 
\begin{center}
\includegraphics[width=6cm,height=8.5cm,angle=270]{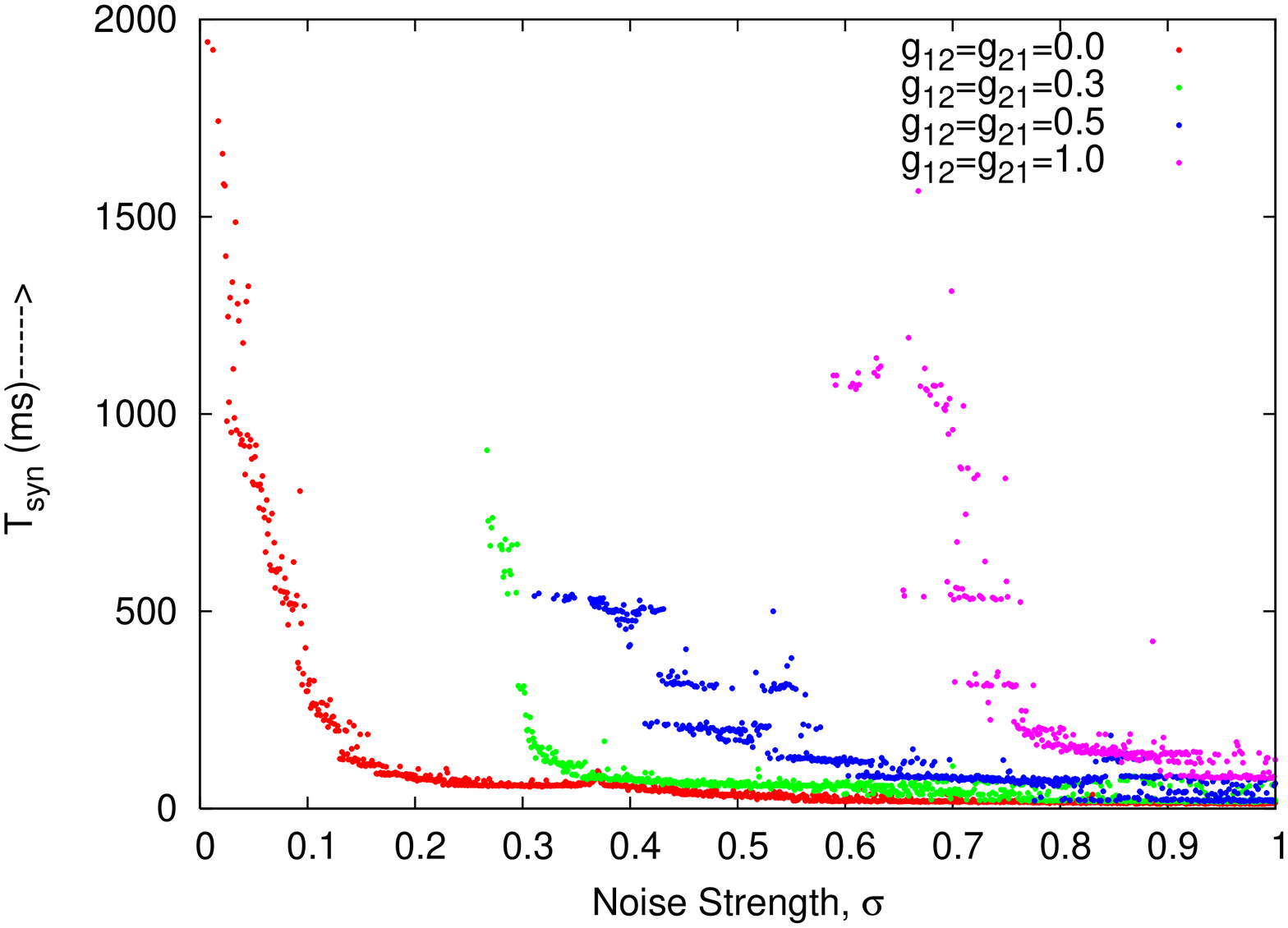}\\
\end{center}
%\vspace*{1cm}
{\bf Fig. 20} (color online)

\end{document}